\documentclass[aps,floats,nofootinbib,amssymb,preprint,superscriptaddress]{revtex4}
\usepackage{epsf,epsfig}

\usepackage{graphicx}
\usepackage{amssymb}

\newcommand{\be}{\begin{equation}}
\newcommand{\ee}{\end{equation}}
\newcommand{\bea}{\begin{eqnarray}}
\newcommand{\eea}{\end{eqnarray}}
\newcommand{\nn}{\nonumber}

\renewcommand{\check}{({\bf This statement needs to be checked!!})}

\newcommand{\half}{\frac{1}{2}}

\newcommand{\ba}{\begin{array}}
\newcommand{\ea}{\end{array}}
\newcommand{\bi}{\begin{itemize}}
\newcommand{\ei}{\end{itemize}}
\newcommand{\ben}{\begin{enumerate}}
\newcommand{\een}{\end{enumerate}}

\newcommand{\N}{\chi^0}
\newcommand{\C}{\chi^\pm}

\newcommand{\mb}[1]{\mbox{\boldmath $#1$}}

\preprint{
\hbox to \hsize{
\hfill$\vcenter{\hbox{\bf MADPH-07-1494}
                \hbox{August 2007}}$}
}

\begin{document}
\title{\vspace*{.5in}
High energy neutrinos from neutralino annihilations in the Sun}

\author{Vernon~Barger}
\email{barger@pheno.physics.wisc.edu}
\affiliation{Department of Physics, University of Wisconsin, 1150 University
Avenue, Madison, Wisconsin 53706 USA}

\author{Wai-Yee~Keung}
\email{keung@uic.edu}
\affiliation{Physics Department, University of Illinois at Chicago, 
Illinois 60607--7059 USA}

\author{Gabe~Shaughnessy}
\email{gshau@hep.wisc.edu}
 
\author{Adam~Tregre}
\email{tregre@hep.wisc.edu}
 
\affiliation{Department of Physics, University of Wisconsin, 1150 University
Avenue, Madison, Wisconsin 53706 USA}

\begin{abstract}
Neutralino annihilations in the Sun to weak boson and top quark pairs lead to high-energy neutrinos that can be detected by the IceCube and KM3 experiments in the search for neutralino dark matter.  We calculate the neutrino signals from real and virtual $WW, ZZ, Zh$, and $t \bar t$ production and decays, accounting for the spin-dependences of the matrix elements, which can have important influences on the neutrino energy spectra.  We take into account neutrino propagation including neutrino oscillations, matter-resonance, absorption, and $\nu_\tau$ regeneration effects in the Sun and evaluate the neutrino flux at the Earth.  We concentrate on the compelling Focus Point (FP) region of the supergravity model that reproduces the observed dark matter relic density.  For the FP region, the lightest neutralino has a large bino-higgsino mixture that leads to a high neutrino flux and the spin-dependent neutralino capture rate in the Sun is enhanced by $10^3$ over the spin-independent rate.  For the standard estimate of neutralino captures, the muon signal rates in IceCube are identifiable over the atmospheric neutrino background for neutralino masses above $M_Z$ up to 400 GeV.

\end{abstract}
\thispagestyle{empty}

\maketitle

\section{Introduction}
A stable Weakly Interacting Massive Particle (WIMP) of mass of order 100 GeV that was produced in the early Universe, thermalized, and froze-out due to the Hubble expansion provides a natural explanation for observed density of dark matter today~\cite{Kolb:1990vq}.  A well motivated dark matter (DM) candidate is the Lightest Stable Particle (LSP) of Supersymmetry (SUSY) with R-parity conservation~\cite{Drees:2004jm,Baer:2006rs,Binetruy:2006ad}. The LSP is nominally the lightest neutralino (denoted by $\N_1$), a neutral spin-1/2 particle that is a linear combination of gauginos (spin-1/2 SUSY companions of the Spin-1 Standard Model bino and Wino) and higgsinos (spin-1/2 companions of two spin-0 Higgs bosons)~\cite{Haber:1984rc}.  The minimal supergravity model (mSUGRA) has minimal SUSY particle content with the exact supersymmetry broken by gravity~\cite{Chamseddine:1982jx} .  With gauge and Yukawa coupling at the Grand Unified Scale (GUT)~\cite{Barger:1992ac,Barger:1993gh,Kane:1993td} the mSUGRA model predictions are given in terms of a small number of parameters~\cite{Ellis:2003cw,Abel:2000vs}.  The consequences of mSUGRA for neutralino dark matter detection have been the subject of numerous phenomenological studies that have considered the predictions for forthcoming experiments; some recent surveys are given in Refs.~\cite{DMSAG,Jungman:1995df,Strumia:2006db,Mena:2007ty,Baltz:2006fm,Hooper:2003ka,Arnowitt:2006jq,Roszkowski:2007va}.  There are three complementary experimental approaches: direct detection via nuclear recoils from WIMP scattering, indirect detection via astrophysics experiments wherein WIMP annihilations give neutrino, gamma ray, positron, and antideuteron signals, and collider experiments where the supersymmetric particles undergo cascade decays to final states with two LSPs that give missing energy in the events.

Our focus is on the signals in neutrino telescopes that should result from the annihilations of neutralinos that have been gravitationally captured by the Sun~\cite{Jungman:1994jr,Barger:2001ur}.  Neutrino telescopes are poised to search for the high energy neutrinos of this origin~\cite{Halzen:2005ar}.  The IceCube experiment at the South Pole is underway and expects to have about 50,000 events from atmospheric neutrinos in the near future~\cite{GonzalezGarcia:2005xw,icecube:2001aa}.  The KM3 detector in the Mediterranean Sea is currently being built~\cite{Sapienza:2005tz,Aguilar:2006rm,Resvanis:2006sb,nestor:2007xx,Antares:2007xx,KM3:2007xx}.  The IceCube detector has a neutrino energy threshold of 50 GeV and the KM3 detector is expected to have sensitivity to neutrinos above a threshold of 10 GeV.  These experiments are expected to have the capability to find or limit signals from neutralino DM annihilations in the Sun~\cite{Barger:2001ur}.  It is therefore of particular interest to refine the characteristic features imprinted on the neutrino energy spectra of this origin.  

We make substantial improvements on previous theoretical studies~\cite{Cirelli:2005gh} by including the full spin-dependence of the matrix elements from real and virtual $WW, ZZ, Zh$, and $t \bar t$ production and subsequent decays to neutrinos, by noting the importance of the spin-dependent capture rate in the Sun, and by concentrating on the region of mSUGRA parameter space known as the Hyperbolic Branch (HB)~\cite{Chan:1997bi} or Focus Point (FP)~\cite{Feng:1999mn} region, where the neutralino annihilations have the highest rates.  (Hereafter we denote this region simply by FP).  The above mentioned annihilation processes lead to the high energy neutrinos of interest for neutrino telescopes.   We propagate the neutrinos through the Sun, taking into account neutrino oscillations, matter resonance, absorption, and $\nu_\tau$ regeneration effects, and give the resulting muon-neutrino energy spectra at Earth and the muon energy spectra in km$^2$ area detectors.  

Neutralinos are gravitationally trapped and accumulate at the center of the Sun where capture and annihilation rates equilibrate over the lifetime of the Sun.  The solar capture rate of neutralinos in the galactic halo is approximately given by \cite{Gould:1992xx}
\be
C_{\odot}= 3.4\times 10^{20} {\rm s}^{-1} {\rho_{local}\over 0.3 \text{ GeV/cm}^3} \left({270\text{ km/s}\over v_{local}}\right)^3\left({\sigma_{SD}^H+\sigma_{SI}^H + 0.07 \sigma_{SI}^{He}\over 10^{-6}\text{ pb}}\right)\left({100\text{ GeV}\over m_{\N_1}}\right)^2,
\label{eq:caprate}
\ee
where $\rho_{local}$ and $v_{local}$ are the local density and velocity of relic dark matter, respectively.  The average density is taken to be $\rho_{local} \approx 0.3  \text{ GeV/cm}^3$, but may be enhanced due to caustics in the galactic plane~\cite{Sikivie:2001fg}.  The capture rate is highly dependent on the strength of the neutralino interactions with matter~\cite{Jungman:1995df,Bertin:2002ky}.  The spin-independent (SI) and spin-dependent (SD) scattering rates determine how strongly the Sun slows and captures the neutralinos and are limited by present direct detection experiments.  The factor of $0.07$ before $\sigma_{SI}^{He}$ comes from the relative abundance of helium and hydrogen in the Sun, as well as other dynamical and form factor effects~\cite{Halzen:2005ar}.  For a discussion of the present limits on the SI and SD cross sections from direct detection experiments see Ref. \cite{DMSAG}.

As neutralinos accumulate in the solar core, their annihilations deplete the population.  The competition between neutralino capture and annihilation can be expressed through the annihilation rate, $\Gamma$, in the solar core by
\be
\Gamma = \half C_{\odot} \tanh^2\left(\sqrt{C_{\odot} A_{\odot}}t_{Sun}\right),
\ee
where $A_{\odot}={\langle \sigma v\rangle\over V}$ is the annihilation rate times relative velocity per unit volume and $t_{Sun}=4.5$ Gy is the age of the Sun. When $\sqrt{C_{\odot} A_{\odot}}t_{Sun}\gg1$, the capture and annihilation processes are in equilibrium, which is expected to be a good approximation for the Sun.  Thus the annihilation rate is equivalent to half the solar capture rate today since two neutralinos create one annihilation event.

From the thermalization of the captured neutralinos, it has been estimated that the neutralinos are concentrated with a core region of very small radius, of order 0.01$R_{Sun}$~\cite{Cirelli:2005gh}, so we can approximate the source of the neutrinos from annihilations as arising from the center of the Sun.  We propagate the neutrinos from the solar center to the surface, taking into account neutrino oscillations with MSW effects~\cite{PhysRevD.17.2369,PhysRevD.22.2718,Mikheev:1986gs}, as well as absorption and reinjection of neutrinos resulting from neutral current (NC) and charged current (CC) interactions.  Since neutralino annihilation is flavor blind\footnote{Strictly speaking, secondary neutrinos from the $\tau$-leptons and $b$-quarks from $W$ and $Z$ decays break this flavor democracy but these contributions result in softer neutrino energy spectra that are less likely to pass the experimental acceptance cuts.}, the relative populations of different neutrino species are the same at the center of the Sun.  MSW effects that change the neutrino flavors as they propagate through the Sun are small in the present application because the injection spectra are uniformly populated among flavors (i.e. the commutator which describes the evolution of the neutrino states as they propagate through the Sun vanishes, yielding a steady state solution).  In addition, MSW effects are smaller at energies above order 100 GeV that are of present interest\footnote{A concurrent study in Ref.~\cite{Lehnert:2007fv} analytically considers the MSW effects of low energy neutrinos from dark matter with mass up to 100 GeV.}.  In the results that we present here, we take $\theta_{13} = 0$, but we do not anticipate substantial changes for nonzero $\theta_{13}$ over its experimentally allowed range\footnote{Its experimentally allowed range is $\sin^2 2\theta_{13}< 0.19$~\cite{Yao:2006px}.}.  The analysis of Ref.~\cite{Cirelli:2005gh} found that a value of $\sin^2 2\theta_{13} = 0.04$ gave little change in the overall neutrino spectra in the high energy region.  

Absorption and reinjection effects due to NC and CC interactions are important.  The effects from NC interactions are neutrino flavor blind and tend to soften the initial spectra; when a neutrino of energy $E_\nu$ is absorbed, a neutrino of the same flavor but lower energy $E'_\nu <E_\nu$ is reinjected into the flux. The CC absorption is also flavor blind, but the resulting reinjected spectra are not, due to the regenerated  $\bar \nu_e$ and $\bar \nu_\mu$ from the leptonic decays of $\tau^-$ leptons.  A $\tau^-$ decays promptly 100\% to $\nu_\tau$ and 18\% to each of $\bar \nu_e$ and $\bar \nu_\mu$\footnote{The $\mu^\pm$ leptons can also inject $\bar \nu_e$ and $\nu_e$, but are stopped by ionizing radiation before they decay, resulting in very soft neutrinos~\cite{Crotty:2002mv}.}.  Although the $\nu_\tau$ number population remains the same, the $\nu_\tau$ energy spectrum is softened by the interactions.  Additionally, other neutrino flavors receive contributions from the $\nu_\tau$ CC interaction.  This asymmetry in the population of neutrino species introduces small, but nonetheless non-negligible vacuum oscillation and NC and CC effects.  Once the neutrinos emerge from the Sun, they propagate as matter eigenstates to the Earth where they can be detected in neutrino telescopes by the high energy muons and electrons produced by the CC interactions in the surrounding ice or water. 

The above effects were included in a number of works~\cite{GonzalezGarcia:2005xw,Cirelli:2005gh,Strumia:2006db}.  In Ref. \cite{Cirelli:2005gh} the reconstruction of a general dark matter candidate was studied.  However, since the spin of the DM particle was not specified, the spin-dependence of the amplitudes could not be taken into account.  The intent to treat the DM annihilations in a model independent way is laudable, but this approach can have serious shortcomings.  For instance, DM annihilation in the static limit to the two body neutrino-antineutrino process is absent in SUSY models, due to helicity suppression.  Additionally, processes such as annihilations to weak bosons exhibit a striking helicity dependence in the neutrino spectra from their decays.  The spectra of neutrinos of energy ${\cal O}(100\text{ GeV})$ can be substantially altered by the spin dependence of the matrix elements.

In Section \ref{sect:susyparm}, we review the regions of mSUGRA parameter space that can account for the observed relic density of neutralinos~\cite{Spergel:2006hy}.  We concentrate on the FP region which solves a variety of problems with SUSY phenomenology~\cite{Feng:1999zg,Feng:2000bp,Baer:2005ky}.  In this region of parameter space, the lightest neutralino is a bino-higgsino mixture and the neutralino annihilation rates are large.  In Section \ref{sect:prod}, we discuss the neutrino production modes that occur through neutralino annihilation in the present epoch.  In Section \ref{sect:spect}, we calculate the neutrino energy spectra for each contributing process. We give analytic formulae for the  neutrino spectra in the static limit where the intermediate state heavy particles are on-shell and show that they closely approximate the matrix elements obtained from SMADGRAPH~\cite{Alwall:2007st,Maltoni:2002qb,Cho:2006sx,Stelzer:1994ta,Murayama:1992gi}.  Propagation of the high energy neutrinos from the center of the Sun to the km$^2$ area detectors on Earth are discussed in Section \ref{sect:nuprop}.  The prospects of detecting neutrinos from neutralino annihilation are presented in Section \ref{sect:nudetect}.  We provide our conclusions in Section \ref{sect:conc}.

\section{SUSY Parameter space}
\label{sect:susyparm}

In our illustrations we implicitly assume the model of minimal supergravity (mSUGRA).  In this model the scalar, gaugino and trilinear masses unify at the grand unification (GUT) scale.  The predictions are determined by the five parameters~\cite{Abel:2000vs}
\be
m_0, m_{1/2}, A_0, \tan \beta, sign(\mu)
\ee 
Here $m_0,m_{1/2}$ and $A_0$ are  the common scalar, gaugino and trilinear masses at the GUT scale.  The parameter $\tan \beta$ is the ratio vacuum expectation values of the up-type and down-type neutral Higgs fields and $\mu$ is the supersymmetry conserving higgsino mass parameter at the electroweak scale.  The absence of a Higgs signal from LEP and the relic density determination from WMAP~\cite{Spergel:2006hy} limit the mSUGRA parameter space to four distinct regions:

\ben
\item The focus point with large values of $m_0$ and relatively small values of $\mu$ can naturally yield the observed relic density~\cite{Baer:2005ky,Chan:1997bi,Feng:1999mn,Feng:1999zg,Feng:2000bp,Baer:1995nq,Baer:1995va}.  This region provides a decoupling solution to the SUSY flavor and CP problems, suppress proton decay rates, and has low fine tuning.  The FP has large sfermion masses and mixed higgsino-bino dark matter and can be suitable for sparticle mass measurements at the LHC~\cite{Baer:2007ya} .  The annihilation modes of the lightest neutralino are predominantly to weak bosons and top quark pairs due to the large higgsino-bino mixture of the lightest neutralino.  

\item The slepton co-annihilation region has low $m_0$ and the lightest slepton (stau) mass is almost degenerate in mass with the lightest neutralino mass~\cite{Ellis:1998kh,Gomez:1999dk,Lahanas:1999uy,Baer:2002fv}.  The resulting co-annihilation can drop the neutralino relic density into the observed range.

\item The bulk region is characterized by low $m_0$ and $m_{1/2}$~\cite{Baer:1995nc,Barger:1997kb}.  In this region, neutralino annihilation is dominated by the $t$-channel exchange of light scalar fermions.  However, this region is now disfavored by the limits from LEP on the lightest chargino and sleptons ~\cite{Ellis:2003cw,Baer:2003yh}.

\item The Higgs funnel region is characterized by large values of $\tan \beta$ where the lightest neutralino annihilates via a light $h$ or broad $A$ resonance~\cite{Drees:1992am,Baer:1995nc,Baer:1997ai,Baer:2000jj,Ellis:2001ms,Roszkowski:2001sb,Djouadi:2001yk,Lahanas:2001yr}.  
\een

\begin{figure}[htbp]
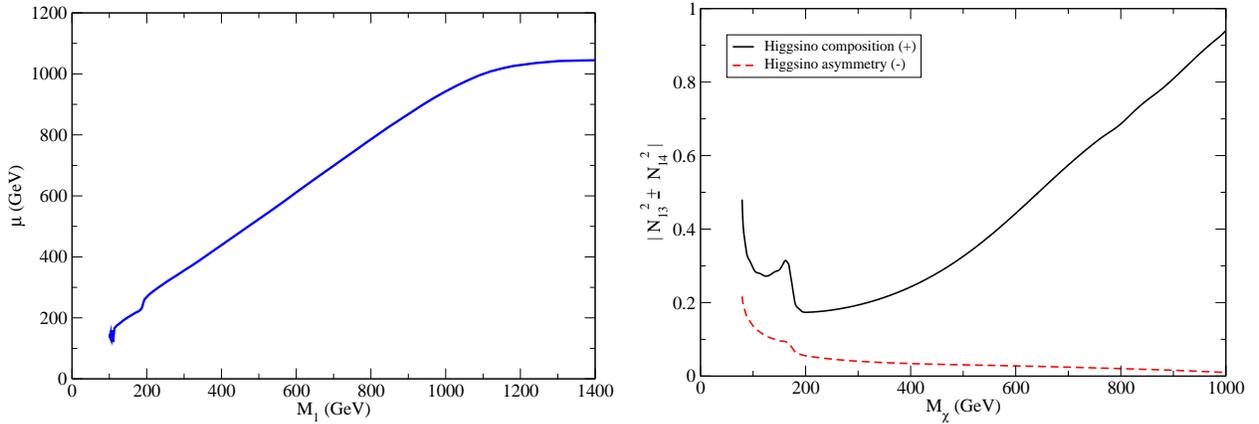

   \begin{center}
      \includegraphics[width=0.49\textwidth]{plots/M1-mu.eps}\hspace{2mm}
      \includegraphics[width=0.49\textwidth]{plots/fp-higgsino.eps}
      \caption{The central FP region in $\mu$ and $M_1$ (left panel) with $m_t=174.3$ GeV and $\tan \beta = 50$ as given in~\cite{Baer:2005ky}. As the mass of the  lightest neutralino increases, it becomes largely higgsino shown by the solid, black curve (right panel).  However, the higgsino asymmetry, an important quantity for the $Z\N_1\N_1$ coupling, decreases as shown by the red, dashed curve.}
      \label{fig:fpscan}
   \end{center}
\end{figure}

Throughout the remainder of this paper, we adopt the low energy determination of the FP region given in Ref. \cite{Baer:2005ky}, which is preferred for $b-\tau$ Yukawa unification~\cite{Barger:1993vu,Bardeen:1993rv}.  This parameterization adopts $\tan \beta = 50$ and $m_t=174.3$ GeV.  The central values of this narrow wedge are reproduced in Fig.~\ref{fig:fpscan} in the space of the parameters $\mu$ and $M_1$ at the electroweak scale, where $M_1$ is the $U(1)$ gaugino mass parameter.  The narrow allowed band lies just above the region where radiative electroweak symmetry breaking is not allowed.  The sfermions and heavy Higgs bosons decouple, so $\mu$ and $M_1$ are the critical parameters for FP phenomenology of the light gauginos.  The mass matrix of the neutralinos in the $(\tilde B^0, \tilde W^0, \tilde H^0_d, \tilde H^0_u)$ basis given in terms of these parameters is
\bea
{\cal M}_{\N} = \left( \begin{array} {c c c c }
	M_1 	&	0&	{-g_1 v_d/ 2}&	{g_1 v_u / 2}\\
	0 	&M_2&	{g_2 v_d / 2}&	{-g_2 v_u / 2}\\
	{-g_1 v_d / 2} 	&	{g_2 v_d / 2}&	0&	-\mu\\
	{g_1 v_u / 2} 	&	{-g_2 v_u / 2}&	-\mu& 0\\
	\end{array} \right),
	\label{eq:neutmass}
\eea
where $g_1$ and $g_2$ are the $U(1)$ and $SU(2)$ gauge couplings, respectively.  Gaugino mass unification restricts the $SU(2)$ gaugino mass $M_2 = {3 g_2^2\over 5 g_1^2} M_1\approx 2 M_1$, and consequently the lightest neutralino has substantially more bino content than wino.  The mass matrix is diagonalized by the rotation matrix $N_{ij}$
\be
{\cal M}_{\N}^D = N^* {\cal M}_{\N} N^T.
\ee
The lightest neutralino can then be expressed in the gaugino-higgsino eigenbasis as
\be
\N_1 = N_{11} \tilde B+N_{12} \tilde W+N_{13} \tilde H_d+N_{14} \tilde H_u.
\ee
Using these methods, we numerically diagonalize Eq. (\ref{eq:neutmass}) and determine the mass and composition of the lighest neutralino.  The composition of the lightest neutralino at some representative points in the FP region are given in Table \ref{tab:illustpts}.

\begin{table}[htdp]
\caption{Representative points in the FP region showing the composition of the lightest neutralino.}
\begin{center}
\begin{tabular}{|c|cccc|}
\hline
$M_{\N_1}$ &$ N_{11}$& $N_{12}$& $N_{13}$& $N_{14}$\\
\hline
90 GeV & 0.80& -0.20&0.49&-0.29\\
110 GeV & 0.83& -0.17&0.45&-0.28\\
200 GeV & 0.90& -0.09&0.34&-0.24\\
400 GeV & 0.87& -0.06&0.37&-0.32\\
700 GeV & 0.65& -0.06&0.55&-0.52\\
1000 GeV & 0.24& -0.04&0.69&-0.68\\
\hline
\end{tabular}
\end{center}
\label{tab:illustpts}
\end{table}%

The mass matrix of the charginos in the $(\tilde W^\pm, \tilde H^\pm)$ basis is 
\bea
{\cal M}_{\C} = \left( \begin{array} {c c }
	M_2 	&	\sqrt 2 M_W \cos \beta\\
	\sqrt 2 M_W \sin \beta	&\mu\\
	\end{array} \right).
	\label{eq:charmass}
\eea
The chargino mass matrix is diagonalized by the rotation matrices $U_{ij}$ and $V_{ij}$
\be
{\cal M}_{\C}^D = U^* {\cal M}_{\C} V^T.
\ee

The relic density of the lightest neutralino is very sensitive to its gaugino-higgsino content.  To accommodate the observed relic density, the lightest neutralino is required to have large bino and higgsino components,  leading to the relation $\mu \approx M_1$ for light neutralinos as shown in Fig.~\ref{fig:fpscan}. As the lightest neutralino mass increases, so do the $\mu$ and $M_1$ values and the neutralino mass increasingly controls the annihilation rate until the $\N_1$ mass cannot increase further as the relic density would increase above the observed value.  This behavior is seen by the plateau in Fig.~\ref{fig:fpscan} at high $M_1$ and $\mu \approx 1$ TeV.  The heaviest that the $\N_1$ can be in the FP region (and still give the correct relic density) corresponds nearly pure higgsino content (i.e. $N_{13}^2+N_{14}^2 \sim 1$, where $N_{13}$ and $N_{14}$ are the higgsino components of the lightest neutralino).  However, since the value of $\mu$ is much larger than the off-block-diagonal elements that are ${\cal O}(M_W)$, the higgsino asymmetry, $|N_{13}^2-N_{14}^2|$, that determines the strength of the $Z\N_1\N_1$ coupling, is very small.

Since $\mu\approx M_1$ throughout most of the FP region, the lightest chargino is dominantly higgsino and the heavier chargino state is dominantly Wino, with a mass that is approximately double that of the lightest chargino.

\section{Neutrino spectra}
\label{sect:prod}

High energy neutrinos ($E_{\nu} \gtrsim 50$ GeV) may be created via neutralino annihilation through a variety of channels.  Since the FP region has a neutralino with large higgsino and bino components, certain diagrams dominate, the most prominent of which are the following:

\ben
\item Annihilation to weak boson pairs, $WW$ or $ZZ$, where one or both weak bosons decay leptonically.

\begin{figure}[h]
\begin{center}
      \includegraphics[height=0.19\textwidth]{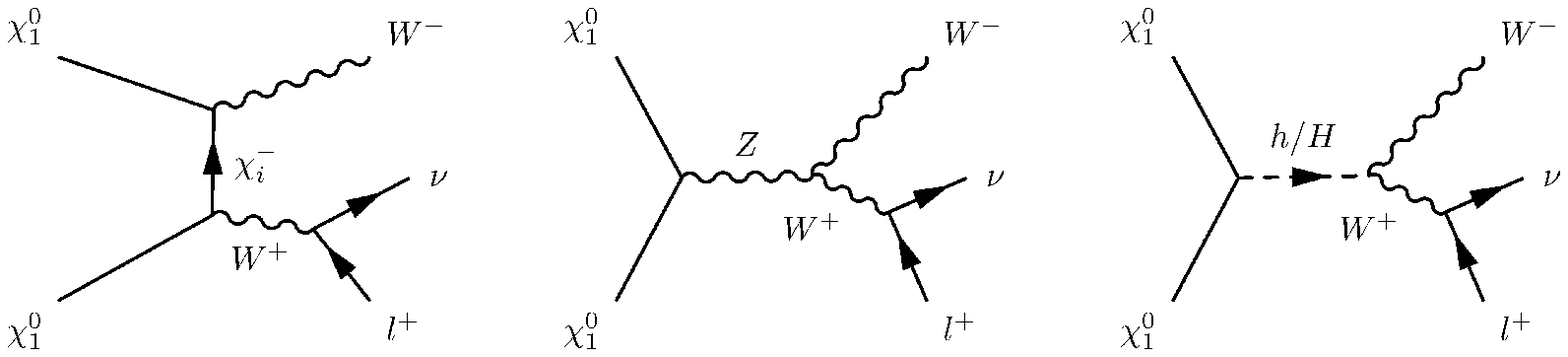}\\
      \includegraphics[height=0.19\textwidth]{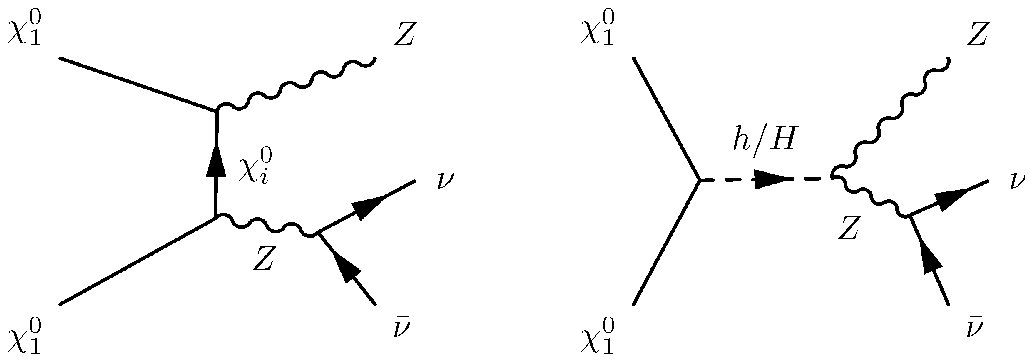}
\caption{Feynman diagrams for the annihilation of the lightest neutralino into $WW$, $ZZ$ where the weak bosons decay into neutrinos.}
\label{fig:anngb}
\end{center}
\end{figure}

The annihilation rate to $WW$ is dominated by the $t$-channel chargino exchange and $s$-channel $Z$ and Higgs boson exchanges shown in the top panels of Fig.~\ref{fig:anngb}.  The chargino diagram can be enhanced by either large higgsino contents of the chargino and neutralino or by a large wino content in the chargino and bino content in the neutralino.  The $ZZ$ and $WW$ diagrams involving $\N_1 \N_1 Z$ vertex are enhanced by the higgsino asymmetry that can be large, as shown in Fig.~\ref{fig:fpscan}.  For large $\tan \beta$, the Higgs boson mass is in the decoupling limit where the light Higgs boson mimics the SM Higgs boson and the masses of the heavy Higgs states, $H^0,A^0$ and $H^\pm$ are nearly degenerate and large.  Therefore, the light Higgs exchange processes require both the higgsino and bino contents to be substantial in the neutralino exchange diagram.  After including all contributing processes, it is found that a strong neutrino signal is expected through the $WW$ and $ZZ$ channels if $M_{\N_1}\gtrsim M_W, M_Z$ and $\mu \lesssim M_1, M_2$.

\begin{figure}[h]
\begin{center}
      \includegraphics[height=0.19\textwidth]{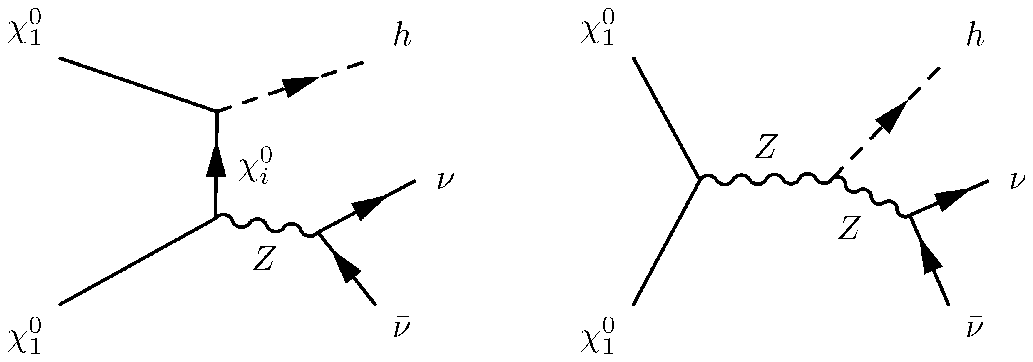}
\caption{Feynman diagrams for the annihilation of the lightest neutralino into $Zh$ where the $Z$-boson and decays into neutrinos.}
\label{fig:annzh}
\end{center}
\end{figure}

\item Annihilation to $Zh$, where the $Z$-boson decays to neutrinos, as shown in Fig.~\ref{fig:annzh}.  This process is enhanced if the higgsino asymmetry is large, due to the $\N_1\N_1 Z$ coupling that occurs in both diagrams.  However, a substantial $t$-channel neutralino exchange amplitude requires a large bino component of the lightest neutralino.  This process is similar to the annihilation to weak bosons except that the emission of the Higgs boson changes the helicity structure of the amplitude which alters the shape of the neutrino energy spectra.

\begin{figure}[h]
\begin{center}
      \includegraphics[height=0.19\textwidth]{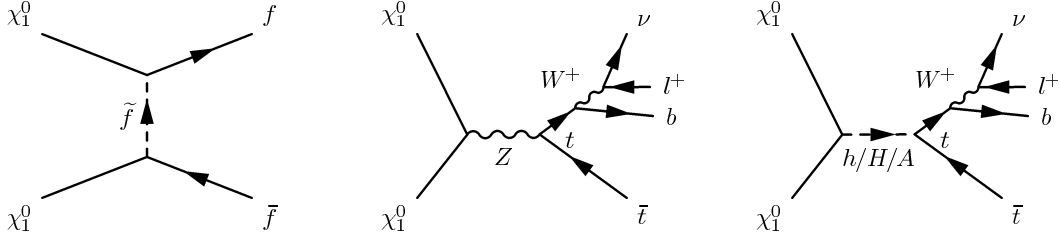}
\caption{Feynman diagrams for the annihilation of the lightest neutralino into $t\bar t$ and decays into neutrinos.  If the scalar fermions decouple, the amplitude for the process in the left panel vanishes.}
\label{fig:anntop}
\end{center}
\end{figure}

\item Annihilation to top quark pairs as shown in Fig.~\ref{fig:anntop}.  Contributions from annihilations to top quarks\footnote{We neglect the QCD corrections~\cite{Moroi:2006fp} to the annihilation to top quark pairs.}, whose primary decay is $t\to \bar b W^+$, with subsequent $W^+ \to \ell^+ \nu$ decay, are dominant if kinematically accessible ($M_{\N_1} > m_t$).  This process can result in a softer neutrino energy spectra compared to direct $W^+W^-$ leptonic decays due to the smaller available phase space for the neutrinos.  The scalar fermion exchange is suppressed due to the large $m_0$ values in the FP region while the $Z$ and Higgs boson exchanges require higgsino asymmetry and higgsino-bino dominance, respectively.  

\item Neutralino annihilation to light quarks and leptons.  Due to the Majorana nature of the neutralino, the annihilation amplitude to light fermions is suppressed by the fermion mass, and direct production of a neutrino-antineutrino pair is therefore negligible\footnote{Note in the analysis in Ref.~\cite{Cirelli:2005gh} of the neutrino spectra from dark matter annihilation it was assumed that the dark matter particle was not Majorana, yielding very clean and distinct neutrino signals from the direct dark matter annihilation to neutrino pairs.  This possibility is not allowed in SUSY models where the neutralino DM candidate is a Majorana particle.}.  If a $\gamma$ or $Z$ boson is emitted, the helicity suppression can be lifted.  The leading contribution to $\N_1\N_1\to q\bar q Z$ via $t$-channel squark exchange is $\alpha^3 m_{\N_1}^6/M_{\tilde q}^8$ for heavy squark masses~\cite{Barger:2006gw}.  In addition to squark exchange, $s$-channel Higgs and $Z$ boson processes contribute when the lightest neutralino is dominantly higgsino.  We do not consider the smaller contributions from these processes.  Moreover, since the neutralino is above the $W$ mass throughout most of the parameter space in the FP region, these modes are negligible\footnote{The neutralino lower mass bound is approximately 80 GeV due to the lower mass limit on the lightest chargino ($M_{\C_1} > 104$ GeV) and the gaugino mass unification assumption.}.  
\een

\begin{table}[htdp]
\caption{Dominant subprocesses for neutrino annihilation to final states that decay to neutrinos.  The $\N_1$ components are given that maximize the subprocess contributions.  Note that since $M_1\sim \mu$, the $\N_{1,2,3}$ states have a large higgsino and/or bino fraction and relatively similar masses.  The light chargino is mainly higgsino while the heavier chargino is mainly Wino.  The top quark Yukawa coupling is denoted by $Y_t$.  The higgsino, bino and Wino contents are denoted by $\widetilde H$, $\widetilde B,$ and $\widetilde W$, respectively. The $\widetilde H$ asymmetry is given by $|N_{13}^2-N_{14}^2|$ and is shown in Fig. \ref{fig:fpscan}.}
\begin{center}
\begin{tabular}{|c|cc|}
\hline
$\N_1\N_1\to WW$	& Relevant $\N_1$ components	& Comments\\
\hline
$t$-channel $\C_1$	& $\widetilde H$	& $\C_1$ dominantly $\widetilde H$	\\
$t$-channel $\C_2$	& $\widetilde W$	&	$\C_2$ dominantly $\widetilde W$ \\
$s$-channel $Z$	& $\widetilde H$ &enhanced by large $\widetilde H$ asymmetry \\
$s$-channel $h$	& $\widetilde H$ \& $\widetilde B$ &\\
\hline
$\N_1\N_1\to ZZ$	&	& \\
\hline
$t$-channel $\N_i$	& $\widetilde H$	&$\N_{4}$ suppressed by small $\widetilde H$ fraction\\
$s$-channel $h$	& $\widetilde H$ \& $\widetilde B$ &\\
\hline
$\N_1\N_1\to Zh$	&	& \\
\hline
$t$-channel $\N_i$	& $\widetilde H$ \& ($\widetilde B$ or $\widetilde W$)	&$\N_{4}$ suppressed by small $\widetilde H$ fraction\\
$s$-channel $Z$	& $\widetilde H$ &enhanced by large $\widetilde H$ asymmetry \\
\hline
$\N_1\N_1\to t\bar t$	&	& \\
\hline
$t$-channel $\tilde t$	& --	&Suppressed by $M_{\N_1}^2 / M_{\tilde f}^4$	\\
$s$-channel $Z$	& $\widetilde H$	&	enhanced by large $\widetilde H$ asymmetry \\
$s$-channel $h$	& $\widetilde H$ \& $\widetilde B$& Enhanced by large $Y_t$\\
\hline
\end{tabular}
\end{center}
\label{tab:anncont}
\end{table}%

Table~\ref{tab:anncont} summarizes the dominant subprocess discussed above and gives the corresponding neutralino annihilation modes and the composition of the $\N_1$ that yields a large rate in each case.

The total annihilation cross section must remain relatively fixed in order to provide the observed relic density $\Omega_{\N_1}h^2 \approx {0.1\text{ pb}\over \langle \sigma v\rangle}$, where $\langle \sigma v \rangle$ is the thermally averaged cross section at freeze-out.  The DM density is thus inversely proportional to the total annihilation rate to neutrinos today\footnote{Up to factors that include the $p$-wave terms, which are suppressed in the present epoch where $v\sim10^{-3}c$.}.  

\begin{figure}[htbp]
   \begin{center}
      \includegraphics[width=0.49\textwidth]{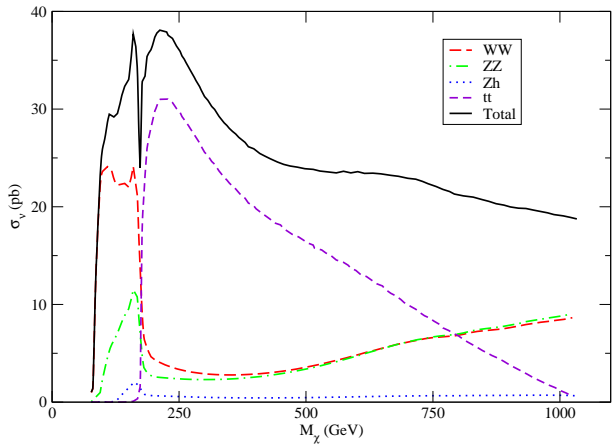}
      \caption{The annihilation cross section of the lightest neutralino to neutrinos via the $WW,ZZ,Zh$ and $t\bar t$ channels in the FP region with $\tan \beta = 50$.  Note that when the top quark threshold is crossed, the cross section for $t\bar t$ production abruptly increases, and the cross sections of all other modes correspondingly decrease in order to give the same $\sigma_{tot}$, as is required to maintain the observed relic density.}
      \label{fig:fpannrate}
   \end{center}
\end{figure}

The annihilation cross section to the neutrinos ($\sigma_\nu$) via the $WW,ZZ,Zh$ and $t\bar t$ modes are shown in Fig.~\ref{fig:fpannrate}.  Once the neutralino mass exceeds the top quark mass, annihilation to top-quark pairs dominates, because the Higgs boson coupling to top quarks, $Y_t$, is large.  However, since the $\N_1$ is then dominantly higgsino with large mass, its coupling to the lightest Higgs boson decreases since some bino or wino content is necessary.    Therefore, at TeV scale $\N_1$ mass, the annihilation to $WW$ and $ZZ$ dominates. 

\section{Calculation of Neutrino Flux}
\label{sect:spect}

The high energy neutrino spectra in the FP region are mainly determined by the diagrams in Figs. \ref{fig:anngb}, \ref{fig:annzh} and \ref{fig:anntop}.  We provide analytic results for the neutrino spectra in the cases of on-shell $WW$, $ZZ$, $Zh$, and $t \bar t$ and show that these formulas reproduce a Monte Carlo (MC) integration of the exact matrix elements provided by SMADGRAPH~\cite{Cho:2006sx}, the supersymmetric version of MADGRAPH.  In the analytic calculation the phase space integration is performed by simple partitioning~\cite{Barger:1987nn} and is summarized in Appendix \ref{apx:ps}.  We calculate the spectra for the illustrative points in Table \ref{tab:illustpts}.

\subsection{Annihilation to weak bosons}

We calculate the annihilation process $\N_1\N_1\to W W^* \to W \ell \nu$ including both on-shell and off-shell contributions.  The annihilation process of $\N_1 \N_1 \to W^+ W^-$ is isotropic in the static limit when the $W$-polarizations are not measured.  The amplitude of the annihilation to on-shell $WW$ is given by the $\chi_i^\pm$ exchange with vanishing contributions from the $s$-channel $Z/H$ diagrams in the static limit,
\be
{\mathcal M} \propto \epsilon_{\alpha \beta \mu\nu} \epsilon^*(k_1)^{\alpha}\epsilon^*(k_2)^{\beta}k_1^{\mu}k_2^{\nu} 
\ee
Because the amplitude involves the Levi-Civita antisymmetric tensor, only the two transversely polarized configurations ($++$ and $--$) of $ W^+W^-$ are allowed and the longitudinal mode is absent. Conservation of $CP$ at tree level implies the two polarization channels carry equal weight.

In the rest frame of $W^+$, the neutrino $\nu$ from  the decay process $W^+\to \ell^+\nu$  is described by the normalized angular distribution $ dN /d\cos\theta = \hbox{$3\over8$} (1- h \cos\theta)^2 $ with $h=\pm 1$ for the right (left) handed polarized $W^+$ with respect to the axis defined by the $W^+$ momentum in the $\N_1\N_1$ CM frame. The antineutrino angular distributions can be obtained by a similar formula with $h\leftrightarrow -h$.  

Including the contributions of both helicities of the $W$, we find the normalized neutrino event rate in the $WW$ CM frame to be 
\be
\label{eq:wwnurate}
{1\over N} {dN\over dE_{\nu}} = {3\over 8 \Delta^3} \left[ (E_\nu-E_c)^2 +\Delta^2 \right],
\ee
where  $\beta = (1 - 4M_W^2/s)^{1\over2}$ is the velocity of the $W$ gauge boson and $E_c=\sqrt{s}/4$ is the average energy.  The energy interval $\Delta=E_c \beta$ is such that $E_c-\Delta < E_\nu < E_c+\Delta$.   The center of mass energy of the process is given by  $s\simeq 4 M_{\N_1}^2$ in the static limit.  The antineutrino from $W^-$ has the same energy distribution.  The result of Eq.(\ref{eq:wwnurate}) is in agreement with Ref. ~\cite{Chang:1992tu}.  Note that the differential rates at the kinematic endpoints of the decay are twice that of energy $m_{\N_1}/4$.  In principle, the distribution in $E_\nu$ could be useful to roughly determine the lightest neutralino mass, provided there is little distortion through NC and CC effects and neutrino propagation.  Ignoring the $W$ helicity, the differential cross section is instead given by a flat $E_\nu$ spectrum, as often applied incorrectly for neutralino DM.
\vspace{1.5mm}

\begin{figure}[t]
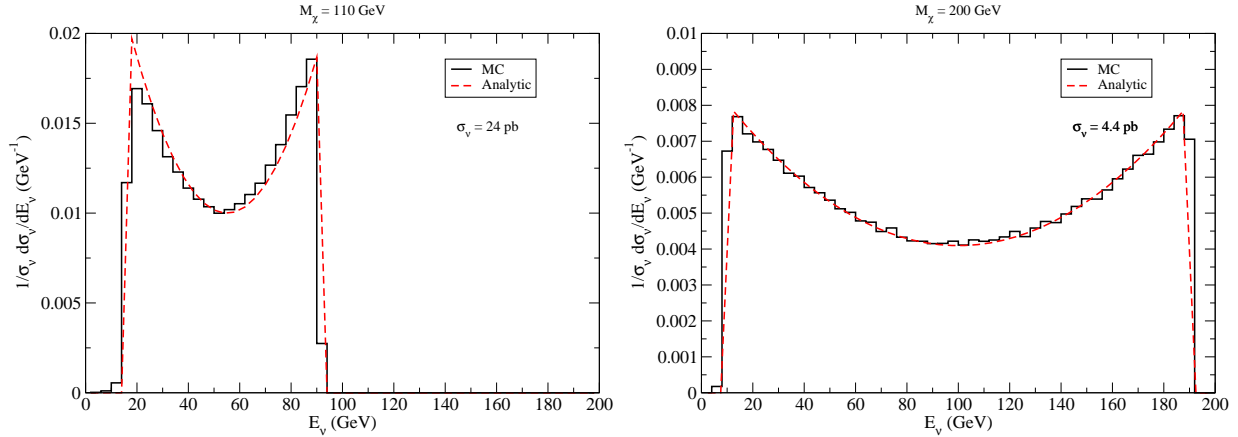

   \begin{center}
      \begin{tabular}{cc}
      \includegraphics[width=0.49\textwidth]{plots/pt02-ww-spectra.eps}
      \includegraphics[width=0.49\textwidth]{plots/pt03-ww-spectra.eps}
      \end{tabular}
      \caption{Neutrino energy spectra for $\N_1\N_1\to WW^*$ and $WW$ where at least one $W$ decays leptonically for $m_{\N_1} =$ 110 and 200 GeV.  Analytic results are given by the dashed curves.}
      \label{fig:wwspect}
   \end{center}
\end{figure}

In Fig.~\ref{fig:wwspect}, we show the analytic and numerical evaluation of the neutrino energy spectra for $\N_1\N_1\to W \ell \nu$ in the FP for $m_{\N_1} =$ 110 and 200 GeV.  The analytic result applies only above $WW$ threshold. Of note is the clear dependence on the polarization and the slight tail on either side of the numerical result, from off-shell decays, in contrast with the sharp cutoff of the analytic result.

The analytic calculation for the neutrino spectra from $\N_1\N_1\to ZZ$ proceeds in a similar way, with $Z$ and $W$ couplings and masses appropriately changed, since the neutrinos have definite helicity.  As the neutralino mass increases, the spectra from $WW$ and $ZZ$ become similar since the W/Z mass difference is relatively small.  Note that even when one gauge boson is off-shell, there is still a rather hard neutrino energy spectra with an appreciable rate.  This is reminiscent of the significant branching fractions of Higgs boson to gauge boson pairs below threshold~\cite{Gunion:1989we}.  Indeed, the neutralino pair in the static limit acts as a pseudoscalar state when coupling to fermions~\cite{Barger:2005ve}.

\subsection{Annihilation to $Zh$} 

Next we analytically calculate the neutrino energy distribution from the static neutralino annihilation, $\chi\chi\to Zh$, $Z\to\nu\bar\nu$.  The primary process $\chi\chi \to Zh$ is isotropic in the static limit.  Only the longitudinal mode of $Z$ is produced in the $\chi\chi$ annihilation because of conservation of angular momentum.

In the rest frame of the longitudinal $Z$, the  neutrino $\nu$ from  the decay process $Z\to \nu \bar\nu$  is described by  the normalized angular distribution with respect to the axis defined by the $Z$ momentum in the $\N_1\N_1$ CM frame, $ dN /d\cos\theta = \hbox{$3\over4$} (1- \cos^2\theta)$.  Then the decay of the $Z$ boson to neutrinos $Z\to \nu\bar\nu$ gives the neutrino spectra
\be
{1\over N} {dN\over dx} =    {3\over4\beta^3 }(1+\beta-x) (x-1+\beta)
\ee                                                                             
where $x=8 M_{\N_1} E_\nu/(4 M_{\N_1}^2+M_Z^2-M_h^2)$ obeys the kinematic range
\be
|x-1|  \le\beta  
\ee

\begin{figure}[t]
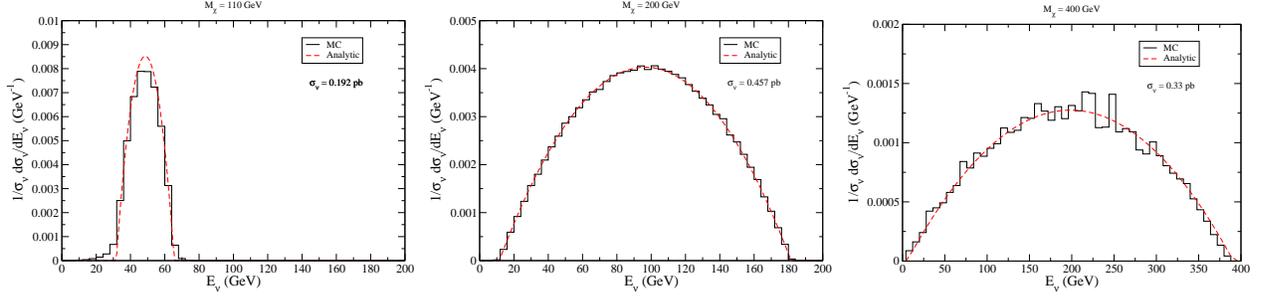

   \begin{center}
      \begin{tabular}{cc}
      \includegraphics[width=0.33\textwidth]{plots/pt02-zh-spectra.eps}
      \includegraphics[width=0.33\textwidth]{plots/pt03-zh-spectra.eps}
      \includegraphics[width=0.33\textwidth]{plots/pt04-zh-spectra.eps}
      \end{tabular}
      \caption{Neutrino energy spectra for $\N_1\N_1\to Zh$ where the Z decays to neutrinos for $m_{\N_1} =$ 110, 200, and 400 GeV.  Analytic results are given by the dashed curves.}
      \label{fig:zhspect}
   \end{center}
\end{figure}

We present these analytic results in Fig.~\ref{fig:zhspect}, along with comparison with our numerical results from SMADGRAPH.  Since the only longitudinal mode of the $Z$ boson contributes, the distribution is very different than that given by the $WW$ or $ZZ$ distribution where only the transverse modes contribute.  

\subsection{Annihilation to heavy fermions}

\begin{figure}[t]
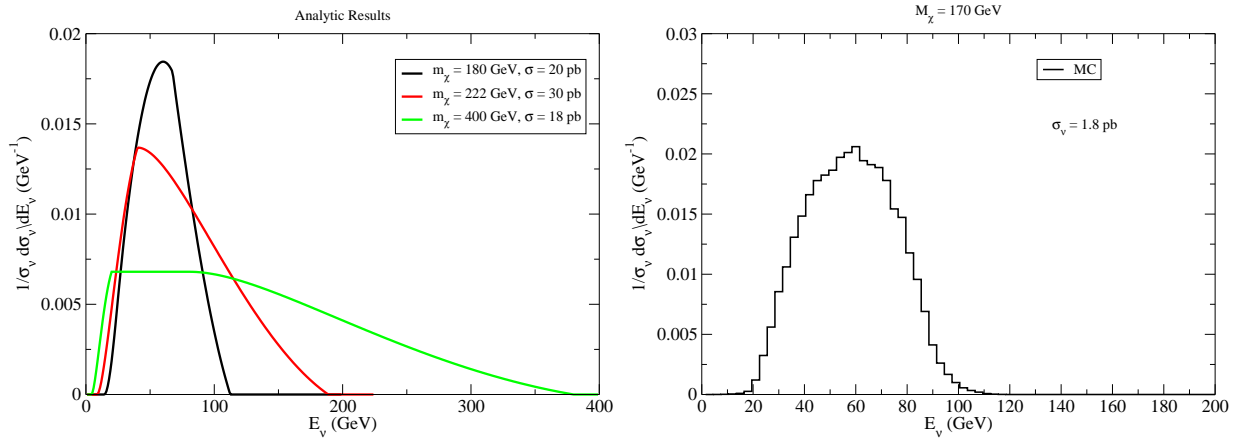

\centering
      \includegraphics[width=0.49\textwidth]{plots/topcut.eps}
      \includegraphics[width=0.49\textwidth]{plots/tt_170.eps}
\caption{(a) MC result for off-shell $t \bar t$ channel, where $M_{\N_1}=\text{ 170 GeV}$. (b) Neutrino energy distributions for various values of $m_\chi$.}
\label{fig:tt}
\end{figure}

The hardest neutrino energies come from annihilation to top quark pairs, when this channel is kinematically accessible.  As the threshold for $t\bar t$ is crossed, the cross section to $ \bar t b \ell^+ \nu_\ell$ jumps dramatically and dominates the overall neutrino rate.  The amplitude of the top quark decay, $t\to b\ell^+\nu$, has the known structure, $ (t\cdot \ell^+)(b\cdot \nu)$, which leads to the neutrino energy spectra in the rest frame of the $t$.  Boosting into the $t \bar t$ center of mass frame, we arrive at the neutrino energy spectra from annihilation to top quarks, assuming a narrow $W$--width, massless $b$-quarks and an average over the $t$-quark polarization
\be {1\over N}{dN\over dE_\nu}={E_t \over\beta D} \left[ F\left(\min({E_\nu\over1-\beta},{E_t\over2})\right)-F\left(\max({E_\nu\over1+\beta},\omega{E_t\over2})\right) \right].
\ee
Here we have defined 
\be
F(y)=(1+2\omega){2y\over E_t}-\hbox{$1\over2$}\left({2y \over E_t}\right)^2 -\omega(1+\omega)\log \left[{2y \over E_t}\right],
\ee
where $\beta$ and $E_t=M_{\N_1}$ are the $t$ velocity and energy, respectively, $\omega = m_W^2/m_t^2$, $D = 1/6 - \omega^2/2 + \omega^3/3$, and the expression applies over the energy range $\left(1 - \beta\right)\omega{E_t \over 2}\leq E_\nu \leq \left(1 + \beta \right) {E_t \over 2}$.  It is interesting to note that the analytic structure of this distribution has two qualitatively different neutrino energy distributions, depending on the neutralino mass.  Specifically, the distribution is divided into three energy regions, the central of which, $|E_\nu-{1\over4}E_t(1+\omega+\beta-\beta\omega)|
\le {1\over4}E_t|1-\omega-\beta-\beta\omega|$, may form a plateau
\bea
 {1\over N}{dN\over dE_\nu}&\propto&\left\{ \begin{array}{ccc} 
 F\left({M_{\N_1} \over 2}\right) - F\left(\omega{M_{\N_1} \over 2}\right), && \text{for }M_{\N_1} > {m_t^2 + m_W^2 \over 2 m_W} \approx 222 \text{ GeV}\\
F\left({E_\nu \over 1 - \beta}\right)-F\left({E_\nu \over 1 + \beta}\right), && \text{for } M_{\N_1} < 222 \text{ GeV}\\
 \end{array}\right. .
 \eea  
These features can clearly be seen in Fig.~\ref{fig:tt}a, which displays the results for the  $E_\nu$ distributions at neutralino masses of $180, 222, \text{ and } 400 \text{ GeV}$.  The high-energy tail of the distribution is described by a simple function of the neutrino energy, namely 
\be
{dN \over d E_\nu} \propto \left({E_\nu \over E_t} - 1\right)\left({ E_\nu \over E_t}-\left(1 + 4 \omega\right)\right) + \omega\left(1+\omega\right)\log { E_\nu \over E_t}.
\ee

We show MC results for the $t^* \bar t\to \bar t b \ell^+ \nu_\ell $ channel in Fig.~\ref{fig:tt}b for the FP with neutralino mass $m_{\N_1}=170$ GeV.

\subsection{Total neutrino energy spectra}

\begin{figure}[t]
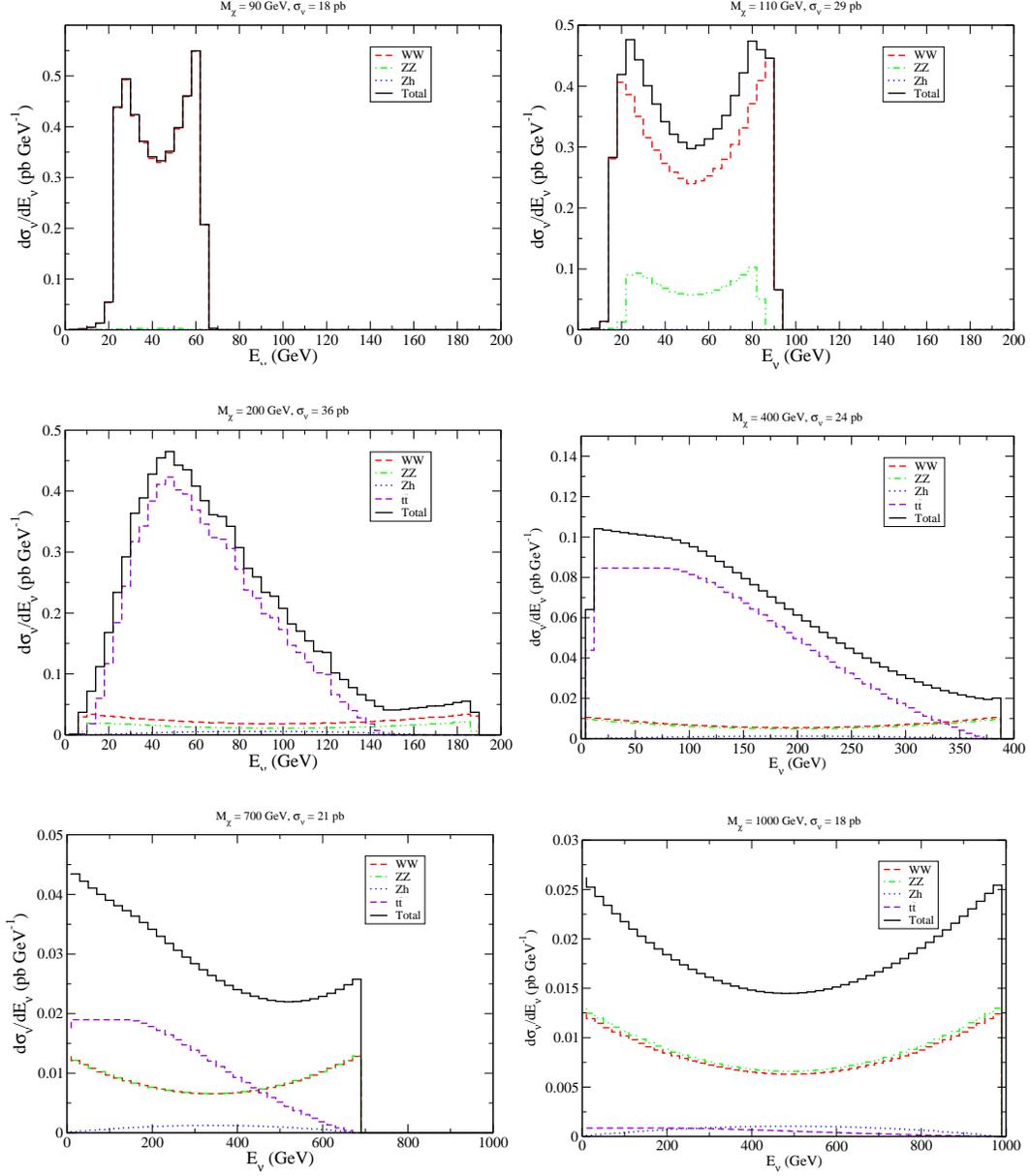

   \begin{center}
      \begin{tabular}{cc}
\vspace{.1in}      \includegraphics[width=0.42\textwidth]{plots/pt01-allspectra.eps}
      \includegraphics[width=0.42\textwidth]{plots/pt02-allspectra.eps}\\
\vspace{.1in}      \includegraphics[width=0.42\textwidth]{plots/pt03-allspectra.eps}
      \includegraphics[width=0.42\textwidth]{plots/pt04-allspect.eps}\\
      \includegraphics[width=0.42\textwidth]{plots/pt06-allspectra.eps}
      \includegraphics[width=0.42\textwidth]{plots/pt05-allspect.eps}
      \end{tabular}
      \caption{Differential cross-sections $d\sigma_\nu/dE_\nu$ at the solar center for representative neutralino masses ranging from 90 GeV to 1000 GeV.  Note that when the neutralino mass exceeds the top mass, the cross section of the $t \bar t$ channel abruptly increases dramatically, and the other cross sections abruptly decrease to maintain the same $\Omega_{DM}$.}
      \label{fig:totspect}
   \end{center}
\end{figure}

We combine the above MC results $WW$, $ZZ$, $Zh$, and $t \bar t$ channels to arrive at the neutrino energy spectra from neutralino annihilation, as shown in Fig.~\ref{fig:totspect} for neutralino masses $m_{\N_1}=90, 110, 200, 400, 700$ GeV, and 1 TeV.  The first two panels illustrate how the $WW$ and $ZZ$ spectra change with $E_\nu$ below the $t\bar t$ threshold.  The on-shell $ZZ$ mode turns on at $m_{\N_1}=M_Z$.   A characteristic double edge feature is apparent in the total spectra at the low and high neutrino energies edges.  However, as the neutralino mass is increased, this double edge becomes less prominent.  Further increasing the neutralino mass includes the dominant top spectra.  In these cases, the upper edge of the spectra in principle would allow a measurement of the lightest neutralino mass.  However, as we will see later, propagation through the Sun can change the neutrino spectra significantly, especially for large neutrino energies, rendering such $\N_1$ mass measurements difficult.  We verify that the $Zh$ signal is naturally suppressed;  it is considerably smaller than the $ZZ$ and $WW$ modes~\cite{Bertin:2002ky,Labonne:2006hk}.

\section{Neutrino propagation}
\label{sect:nuprop}

The neutrino flux at Earth from annihilating neutralinos in the core of the Sun requires a full treatment of the propagation through solar matter and the vacuum between the Sun and Earth.  To retain the information on coherency of the states, we utilize the quantum mechanical evolution equation
\be
\label{eq:prop}
{d {\mb\rho} \over dr} = -i [{\mb H},{\mb\rho}] + \left.{d {\mb\rho}\over dr}\right|_{NC} + \left.{d {\mb\rho}\over dr}\right|_{CC} + \left.{d {\mb\rho}\over dr}\right|_{inj}-\epsilon  [{\mb H}, [{\mb H},{\mb\rho}]],
\ee
where $\mb\rho$ is the complex density matrix in the gauge eigenbasis describing the state of the neutrino as it propagates.  Due to the fact that fast oscillations can only be observed as average values due to finite energy resolution, the epsilon term is included in Eq. (\ref{eq:prop}) for convenience in making the energy average~\cite{Raffelt:1992uj,Strumia:2006db}.  The choice of $\epsilon = 0.01r$ in this decoherence term damps all off-diagonal terms to zero~\cite{Strumia:2006db}.  The Hamiltonian, $\mb H$, includes the vacuum oscillation effects from nonzero mass splitting and the MSW terms:
\be
{\mb H} = {{\mb m}^\dagger {\mb m}\over 2E_\nu}+\sqrt{2} G_F \left[ N_e(r) \delta_{i1}\delta_{j1}-{N_n(r)\over2}\delta_{ij}\right].
\label{eq:ham}
\ee
Here $\mb m$ is the neutrino mass matrix in the gauge eigenstate basis, $E_\nu$ is the neutrino energy, $G_F=1.66\times10^{-5}\text{ GeV}^{-2}$ is the Fermi constant and $N_e(r)$ and $N_n(r)$ are the electron and neutron densities in the Sun~\cite{Bahcall:2004yr}.  The source terms, $\left.{d {\mb\rho}\over dr}\right|_{NC, CC}$ describe the absorption and re-injection of neutrinos caused by Neutral Current (NC) and Charged Current (CC) processes while the injection source term $\left.{d \mb\rho \over dr}\right|_{inj}$ describes the initial spectra injected by neutralino annihilation in the core of the Sun.  

\subsection{Vacuum oscillations and MSW effects}
The ${\mb m}^\dagger {\mb m}$ term in Eq. (\ref{eq:ham}) drives pure vacuum oscillations.  In our illustrations we adopt the vacuum mass-squared difference values of $\Delta m_{23}^2 = 2.5\times 10^{-3}\text{ eV}^2$, $\Delta m_{12}^2 = 8.0\times 10^{-5}\text{ eV}^2$ and mixing angles $\theta_{12} = 35.3^\circ$, $\theta_{23} = 45^\circ$ and $\theta_{13} = 0$.  The matter interactions with the varying solar density suppresses vacuum oscillations.  However, since the injected spectra are uniformly populated among the neutrino species, vacuum oscillations and MSW effects do not change the relative populations, since $[{\mb H},{\mb\rho}] = 0$, making the initial state ${\mb\rho} \sim \mathbb{I}_{3\times3}$ a steady state.

\subsection{Source Terms}
The source terms in the propagation equation account for the absorption and re-injection of neutrinos as the neutrinos propagate through matter.  The CC and NC interactions  absorb higher energy neutrinos and re-inject them with lower energy.  

\subsubsection{Injection spectra}
The source term  $\left.{d {\mb\rho}\over dr}\right|_{inj}$ represents the injected initial neutrino spectra from neutralino annihilation and is given by $\left.{d {\mb\rho}\over dr}\right|_{inj} = \delta(r) \delta_{ij} {1\over N}{d N\over dE_\nu}$.  Here, we make the justified assumption that the neutrino flux is initially injected at the center of the Sun ~\cite{Cirelli:2005gh}. In the following, we illustrate the effect of each term in Eq. (\ref{eq:prop}).  Neutralino annihilation produces neutrinos and antineutrinos with each $\nu$ and $\bar \nu$ flavor having approximately equal probability.

\subsubsection{Neutral Current}
As NC effects are flavor blind, all neutrino species have spectra that are skewed equally towards a lower energy.  The source term is given by
\bea
\left.{d {\mb\rho}\over dr}\right|_{NC} &=&\int^{\infty}_{E_\nu} dE'_{\nu} {d \Gamma_{NC}(E'_{\nu},E_{\nu})\over d E_\nu} {\mb\rho}(E'_{\nu}) \\
&&- {\mb\rho}(E_\nu) \left(N_p(r) \sigma(\nu_l p \to \nu'_l X)+N_n(r) \sigma(\nu_l n \to \nu'_l X)\right).\nn
\eea
The first term represents the re-injection of lowered energy neutrinos with energy $E_\nu$ from an incoming neutrino of energy $E'_\nu$.  The negative term accounts for the absorption of neutrinos of energy $E_\nu$.  

\begin{figure}
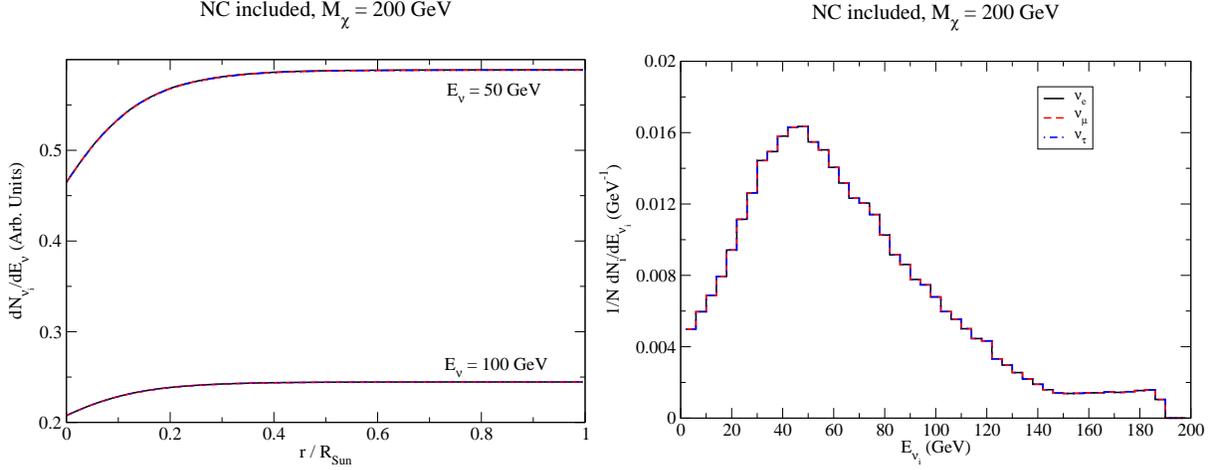

   \begin{center}
      \includegraphics[width=0.47\textwidth]{plots/pt03-rho-r-NC.eps}
      \includegraphics[width=0.49\textwidth]{plots/pt03-spectra-NC.eps}
      \caption{(a) Neutrino spectra vs. radius for $E_\nu = $50 and 100 GeV including the NC source term.  (b)  Neutrino energy spectra at Earth.  Due to the flavor blind NC interactions, the populations of the three neutrino flavors remain unchanged as they propagate.  }
      \label{fig:nc}
   \end{center}
\end{figure}

In Fig.~\ref{fig:nc}a we show the neutrino populations in propagation from the center of the Sun, where annihilation of a $M_{\N_1} = 200$ GeV neutralino occurs, to the solar surface including only the NC effects for $E_\nu=50$ GeV and $E_\nu=100$ GeV.  Due to the flavor diagonal NC interactions, the three neutrino flavors remain distinct as they propagate.  The neutrino spectrum of each flavor is equivalent and given in Fig.~\ref{fig:nc}b.  

\subsubsection{Charged Current}
The CC effects are also flavor blind to the extent that the $\tau$ mass is negligible compared to $E_\nu$.  However, the lepton produced in deep inelastic scattering can be stable (in the case of the electron), can stop before it decays (muon), or decay before radiation losses decrease its energy (tau).  The tau lepton is the dominant source of re-injection of neutrinos.  The channels are
\bea
\nu_\tau N \to \tau^- X &\to& \left\{\begin{array}{c}
\nu_\tau X,\\
\nu_\tau \bar \nu_e e,\\
\nu_\tau \bar \nu_\mu \mu
\end{array}\right.
\eea
where $X$ denotes hadronic decay products.  The source term is
\bea
\label{eq:ccsource}
\left.{d {\mb\rho}\over dr}\right|_{CC} &=& -{\{{\mb\Gamma}_{CC},{\mb\rho}\} \over 2}  + \int {dE_{\nu}^{in} \over E_{\nu}^{in}}[ {\mb\Pi}^\tau \rho_{\tau \tau}\left(E_\nu^{in}\right) {\mb\Gamma}_{CC}^\tau\left( E_\nu^{in}\right) f_{\tau \to \tau}\left( E_\nu/E_\nu^{in}\right)\\\nn
&&+ {\mb\Pi}^{e,\mu} \bar \rho_{\tau \tau}\left(E_\nu^{in}\right) \bar {\mb\Gamma}_{CC}^\tau\left( E_\nu^{in}\right) f_{\bar \tau \to e,\mu}\left( E_\nu/E_\nu^{in}\right) ],
\eea
with a similar term for antineutrinos.  The ${\mb\Pi}^i_{nm} = \delta_{ni}\delta_{im}$ are flavor projection operators where $i=1,2,3$ for the $e,\mu,\tau$ leptons, respectively, and $\mb\Gamma$ is a $3 \times 3$ diagonal matrix which describes the CC interaction rate for a particular flavor, 
\be
{\mb\Gamma}_{\ell \ell} = {\mb\Gamma}_{CC}^\ell = N_p\left(r\right)\sigma\left(\nu_\ell p \to \ell X\right) + N_n\left(r\right)\sigma\left(\nu_\ell n \to \ell X\right).
\ee  
\begin{figure}[t]
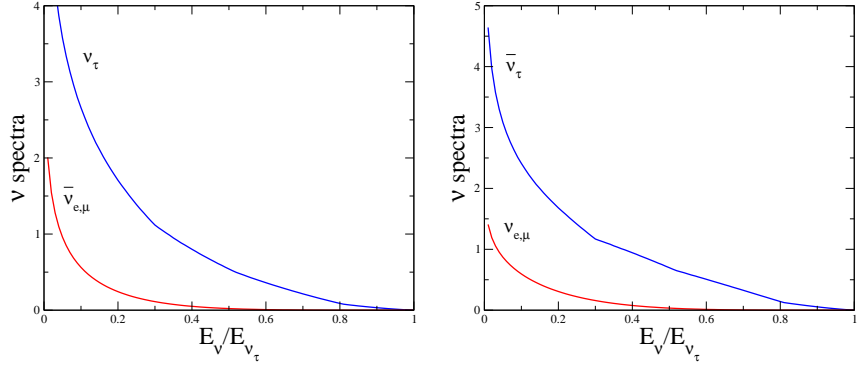

\begin{center}
      \includegraphics[width=0.33\textwidth]{plots/nu_tau_reg.eps}\hspace{3mm}
      \includegraphics[width=0.33\textwidth]{plots/nu_tau_bar_reg.eps}
\caption{Probability density functions, (a) $f_{\tau \to \tau}(u)$ and $f_{\tau \to \bar e, \bar \mu}(u)$ and (b) $f_{\bar\tau \to \bar\tau}(u)$ and $f_{\bar\tau \to  e, \mu}(u)$, where $u=E_\nu/E_\nu^{in}$. }
\label{fig:transfn}
\end{center}
\end{figure}
The functions $f(u)$ in Eq.~(\ref{eq:ccsource}) are the energy distributions of an outgoing neutrino with energy $E_\nu$, given an incoming neutrino with energy $E_\nu^{in}$; they are calculated in Appendix \ref{apx:regen} and shown in Fig.~\ref{fig:transfn}.  Including this $\tau$ regeneration effect, the total $\nu_\tau$ population does not change, but the $\nu_\tau$ energy distribution becomes skewed to lower energy.  The populations $\bar \nu_{e,\mu}$ energy spectra are increased from the leptonic decays of the $\tau^-$ lepton, and likewise for the $\nu_{e,\mu}$ for the $\tau^+$ lepton.

\begin{figure}[t]
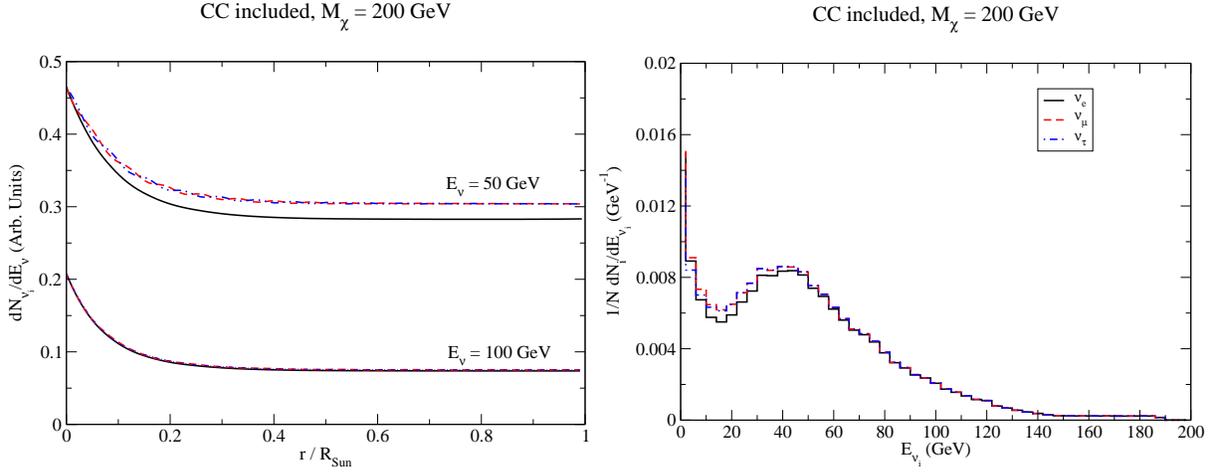

   \begin{center}
      \includegraphics[width=0.47\textwidth]{plots/pt03-rho-r-CC.eps}
      \includegraphics[width=0.49\textwidth]{plots/pt03-spectra-CC.eps}
      \caption{Similar to Fig.~\ref{fig:nc} except that only CC interactions are included.  The re-injection of $\nu_{e,\mu}$ from $\tau$ lepton decays breaks the flavor symmetry among the three neutrino states.}
      \label{fig:cc}
   \end{center}
\end{figure}

In Fig.~\ref{fig:cc} we show the populations of the neutrino similar to those of Fig.~\ref{fig:nc}, but now including only CC interactions.  In this case the $\tau$ decay re-injects $\tau$ neutrinos and occasionally $\bar \nu_{e,\mu}$ neutrinos, breaking the flavor symmetry.  This is clearly seen in the left panel of Fig.~\ref{fig:cc}a for the case of an $E_\nu=50$ GeV neutrino.

\subsection{Propagation through the Sun}

\begin{figure}[ht]
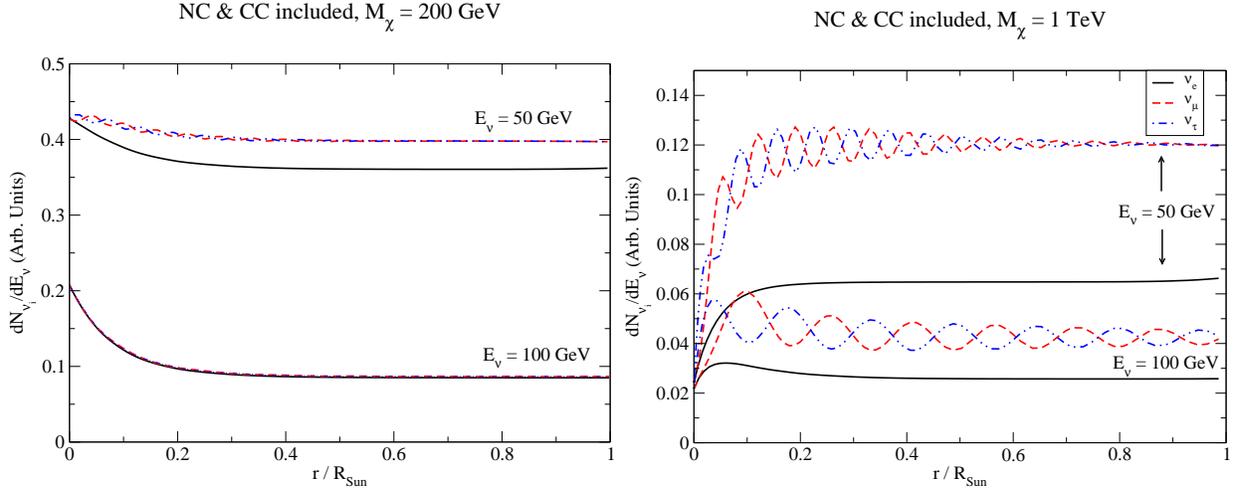

   \begin{center}
      \begin{tabular}{cc}
      \includegraphics[width=0.49\textwidth]{plots/pt03-rho-r.eps}
      \includegraphics[width=0.49\textwidth]{plots/pt05-rho-r.eps}
      \end{tabular}
      \caption{Neutrino spectra after all propagation effects through the Sun are included.  The averaging of oscillations at larger $r/R_\odot$ is achieved by the decoherence term in Eq. (\ref{eq:prop}).}
      \label{fig:allprop}
   \end{center}
\end{figure}

We include all the above effects when propagating the neutrino states to the surface of the Sun.  We illustrate results in Fig.~\ref{fig:allprop} at neutrino energies of 50 and 100 GeV for neutralino masses with values 200 GeV and 1 TeV.  Note that as the neutralino mass increases, the flavor asymmetry increases.  This is due to the larger CC effects at higher energy.

\subsection{Propagation to the Earth}

\begin{figure}
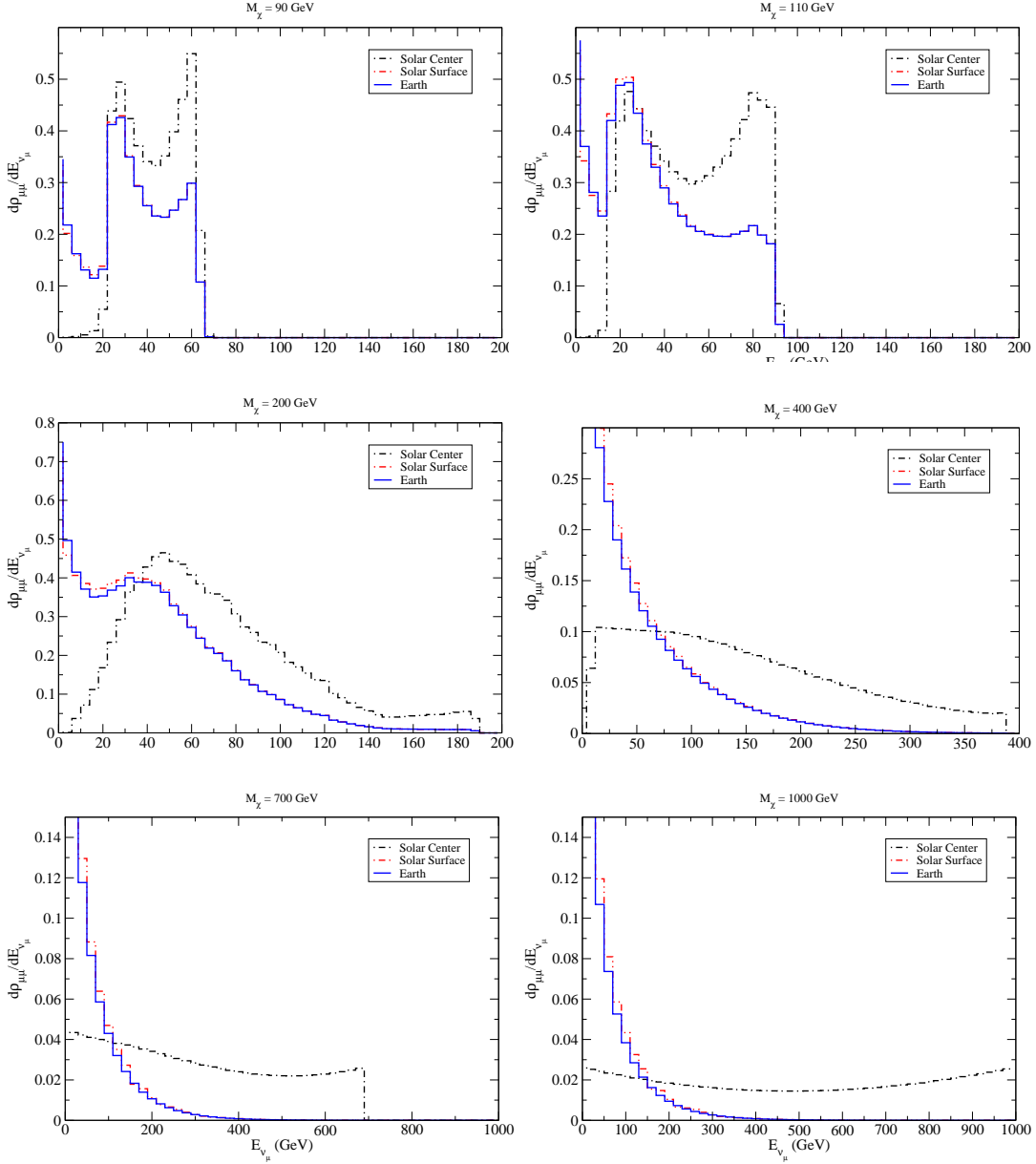

   \begin{center}
      \begin{tabular}{cc}
      \includegraphics[width=0.43\textwidth]{plots/pt01-enu-spect.eps}
      \includegraphics[width=0.43\textwidth]{plots/pt02-enu-spect.eps}\\
      \includegraphics[width=0.43\textwidth]{plots/pt03-enu-spect.eps}
      \includegraphics[width=0.43\textwidth]{plots/pt04-enu-spect.eps}\\
      \includegraphics[width=0.43\textwidth]{plots/pt06-enu-spect.eps}
      \includegraphics[width=0.43\textwidth]{plots/pt05-enu-spect.eps}
      \end{tabular}
      \caption{The neutrino energy spectra at production in the Sun, after propogation to the Sun's surface, and at the Earth's surface.}
      \label{fig:surfEarth}
   \end{center}
\end{figure}

Once propagated to the surface of the Sun, the neutrino states must be propagated in their mass eigenstates to the Earth, which is achieved by rotating the density matrix ${\mb \rho}$ with the unitary MNSP matrix, ${\mb V}$,~\cite{Pontecorvo:1967fh,Gribov:1968kq}
\be
\tilde {\mb \rho}^D = {\mb V}{\mb \rho}_{e,\mu,\tau}{\mb V}^\dagger,
\ee
which diagonalizes the neutrino mass-squared matrix,
\be
{\mb M}^2_\nu = {\mb V} {\mb m}^\dagger {\mb m} {\mb V}^\dagger,
\ee
where ${\mb M}^2$ is the diagonal neutrino mass-squared matrix and ${\mb m}$ is the neutrino mass matrix in the flavor basis.  We use the tri-bimaximal mixing matrix for the rotation~\cite{Harrison:2002er}.  The density matrix in the mass eigenbasis is denoted by $\tilde {\mb \rho}$.  Additionally, the neutrino oscillations are averaged by forcing the off-diagonal elements of $\tilde {\mb \rho}$ to zero in the propagation to Earth.  This averaging could equally well be achieved by including the decoherence term in the propagation equation in Eq. (\ref{eq:prop}).  The neutrino spectra at Earth are quite similar to those at the surface of the Sun, as shown in Fig.~\ref{fig:surfEarth}.

\section{Neutrino Detection}
\label{sect:nudetect}

Once the neutrinos are propagated to Earth, we need to address their rate of detection in a km$^2$ area detector.  The fractional rate of annihilation to neutrinos is given by ${1\over \sigma_{tot}}{d \sigma_\nu \over d E_\nu}$ where $\sigma_{tot}$ includes all SM inclusive modes:  $WW, ZZ, f\bar f, Zh, hh$ and $\sigma_\nu \equiv \sigma_{\N_1 \N_1 \to \nu X}$ includes all the modes discussed above in Section \ref{sect:prod}.  We calculate the total rates using Calchep and can be analytically found in Ref.~\cite{Nihei:2002ij}.

\subsection{Neutrino Flux at Earth}

We illustrate the detection rates using the simulation for IceCube outlined in Ref. \cite{GonzalezGarcia:2005xw}.  The flux of neutrinos from neutralino annihilation at the Earth is given by
\be
{d\Phi_\nu \over dE_\nu} = {1\over \sigma_{tot}}{d \sigma_{\N_1 \N_1 \to \nu X} \over d E_\nu} {1 \over 2} {C_\odot \over 4 \pi R^2},
\ee
where the factor of $\half$ is associated with the fact that two neutralinos produce one annihilation event; here $R=1.49\times10^{11}{\rm m}$ is the Earth-Sun distance.  The parameter $C_\odot$ is the neutralino solar capture rate of Eq. (\ref{eq:caprate}) and is dependent on the neutralino mass, velocity and local density in the galaxy.

\subsection{FP capture rate predictions}

The solar capture rate of neutralinos in the galactic halo is approximately given by Eqn. (\ref{eq:caprate}) and depends on the spin-independent (SI) and spin-dependent (SD) scattering rates\footnote{The present 90\% C.L. limit on the SI cross section from the XENON10 experiment places an upper limit on the scattering rate of $8.8\times10^{-8}$ pb for a WIMP of mass 100 Gev and $4.5\times10^{-8}$ pb for a WIMP of mass 30 GeV~\cite{Angle:2007uj}.  These rates are still higher than the SI rate throughout the FP region, which should be probed in the future by CDMS 2007 with an expected reach of $1 \times 10^{-8}$ pb~\cite{Akerib:2006ri}.  The SD scattering rate off protons, however, is less constrained with a current upper limit of $7\times10^{-2}$ pb from ZEPLIN-II for neutralino masses at 65 GeV~\cite{al.:2007xs}.  Future detectors such as COUPP can greatly improve these limits down to $10^{-2}$ pb with a 2 kg chamber of superheated $CF_3 I$~\cite{Bolte:2006pf}.  However, the SD scattering limits would  still be too weak to constrain the FP region.}.

\begin{figure}[t]
   \begin{center}
      \includegraphics[width=0.49\textwidth]{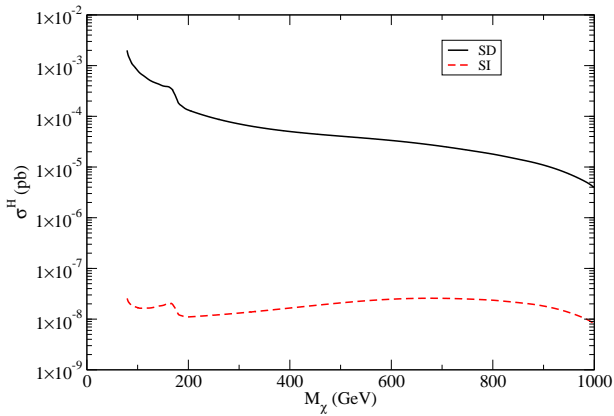}
      \caption{The spin dependent and spin independent scattering cross section of the lightest neutralino with protons. }
      \label{fig:sdhyd}
   \end{center}
\end{figure}

To predict the neutrino flux at Earth, we calculate the spin-dependent cross section for the $Z$ exchange\footnote{Note that the scalar quark exchange is also relevant for the SD scattering rate.  However, since the squarks have large masses in the FP region, they decouple, and squark exchange contributions are not included in the calculation.}
\be
\sigma_{SD}^{\N_1 - p} = {32 \mu_{\chi}^2 J(J+1)\over \pi}   \left[\sum_{q=u,d,s} {\delta_q T_{3}^q \over 2 \sqrt{2}} {g^2\over 4 M_W^2}\left(N_{14}^2-N_{13}^2\right)   \right]^2,
\ee
where $J=1/2$ is the spin of the proton, $\mu_{\chi} = {M_{\N_1} m_p\over M_{\N_1} +m_p}$ is the reduced mass of the neutralino proton system, $T_3^q$ is the isospin of quark $q$ and $\delta_q$ includes the parton distribution function and hadronic matrix element of the proton and are given by the values $ \delta_u = 0.78, \delta_d = -0.48$ and $ \delta_s = -0.15$~\cite{Baer:1997ai,Baer:2003jb}.  The calculated SD and SI cross sections along the FP region are shown in Fig.~\ref{fig:sdhyd}.  The SD cross section varies between ${\cal O}(10^{-3})-{\cal O}(10^{-6})$ pb, while the SI cross section is below $\text{few}\times 10^{-8}$ pb and is consistent with the limit from XENON10 of $4.5\times 10^{-8}$ pb, making the predicted SD cross section dominate over the typical SI rate by a factor of $10^3$.

\subsection{Muon rate at IceCube}
\begin{figure}    
   \begin{center}       
      \begin{tabular}{cc} 
      \includegraphics[width=0.49\textwidth]{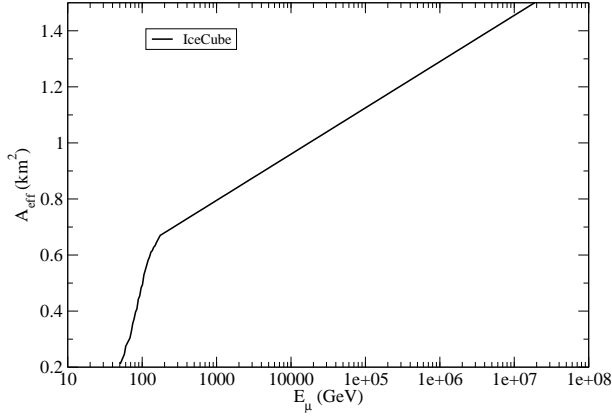}       
      \end{tabular}
      \caption{IceCube effective area for given muon energy.}
      \label{fig:aeff}
   \end{center} 
\end{figure}

Once the neutrino flux is known, the associated total muon rate through a time $T$ in a km$^2$ area detector such as IceCube can be determined by folding the muon production cross section with the neutrino flux
\be
{d N_\mu\over dE_\mu} =   \int_{E_\mu}^{\infty} {d\Phi_\nu \over dE_{\nu_{\mu}}} \left[{d \sigma^p_\nu (E_{\nu_\mu}, E_\mu)\over dE_\mu} \rho_{p}+{d \sigma^n_\nu (E_{\nu_\mu}, E_\mu)\over dE_\mu} \rho_{n}\right] R_\mu(E_\mu) A_{eff}\left(E_\mu\right) dE_{\nu_\mu} + \left(\nu \to \bar \nu\right).
\label{eq:numurate}
\ee
The densities of protons and neutrons near the detector are taken to be $\rho_{p} = {5\over 9}N_A \text{ cm}^{-3}$ and $\rho_{n} = {4\over 9}N_A \text{ cm}^{-3}$, respectively, where $N_A$ is Avagadro's number\footnote{Since the muon range is at most 1 km for a 1 TeV muon, the point of muon production can be assumed to be in ice, rather than the Earth's crust.}.  Muons lose energy according to~\cite{Dutta:2000hh}
\be
{dE \over dx} = -\alpha - \beta E,
\label{eq:muloss}
\ee
which can be used to determine the length of a muon track.  The parameters $\alpha = 2.0\times 10^{-6}$ TeV cm$^2$/g and $\beta = 4.2\times 10^{-6}$ cm$^2$/g describe the loss rate~\cite{GonzalezGarcia:2005xw,Halzen:2005ar,Halzen:2003fi,Dutta:2000hh}.  The muon range,
\be
R_\mu(E_\mu) = {\rho\over \beta} \ln\left[{\alpha + \beta E_\mu\over \alpha+\beta E_\mu^{thr}}\right],
\ee
is the distance a muon propagates through matter of density $\rho$ before its energy drops below the threshold energy, $E_\mu^{thr}$~\cite{Halzen:2003fi}.  We take $E_\mu^{thr} = 50$ GeV, which is optimistic for IceCube and conservative for KM3.  Due to the long muon range, the fiducial volume of the detector can be factored into the range and the cross sectional area of the detector, called the effective area.  The effective area of the detector, $A_{eff}$, is calculated for IceCube following Ref. ~\cite{GonzalezGarcia:2005xw}, and is given in Fig \ref{fig:aeff}.  

The Super-Kamiokande experiment has placed a limit on the flux of muon induced by neutrinos from DM annihilations in the Sun of $\Phi_\nu \lesssim 5\times 10^{-15}\text{ cm}^{-2}\text{s}^{-1}$ for a half-angle of $0-5^\circ$~\cite{Desai:2004pq}.  We calculate the induced muon flux according to 
\be
{d \Phi_\mu \over dE_\mu} = {1\over  A_{eff}(E_\mu)} {dN_\mu \over dE_\mu}
\ee
and find a total muon flux above the Super-K bound for 90 and 110 GeV with $1.3\times 10^{-14}\text{ cm}^{-2}\text{s}^{-1} $ and $7.9\times 10^{-15}\text{ cm}^{-2}\text{s}^{-1}$, respectively.  The other points we study are well below the present bound with  $4.5\times 10^{-16}\text{ cm}^{-2}\text{s}^{-1}$, $2.1\times 10^{-17}\text{ cm}^{-2}\text{s}^{-1}$, $3.2\times 10^{-19}\text{ cm}^{-2}\text{s}^{-1}$ and $4.4\times 10^{-18}\text{ cm}^{-2}\text{s}^{-1}$, for 200, 400, 700 and 1000 GeV neutralino masses along the FP region, respectively.

We apply Eq. (\ref{eq:numurate}) to calculate neutrino signals in IceCube from neutralino annihilations as well as backgrounds from atmospheric neutrinos~\footnote{Backgrounds from solar atmospheric neutrinos also exist, where cosmic rays produce neutrinos in the solar atmosphere~\cite{Fogli:2006jk}.  These backgrounds are expected to be a few per year for a km$^2$ size detector triggering on upward going muons.}.  The differential flux of atmospheric $\nu_\mu$ and $\nu_{\bar \mu}$ neutrinos, $d \Phi_{\nu}/dE_\nu d\cos \theta_z$, are taken from Ref.~\cite{Honda:2006qj}, where $\theta_z$ is the zenith angle.

\begin{figure}
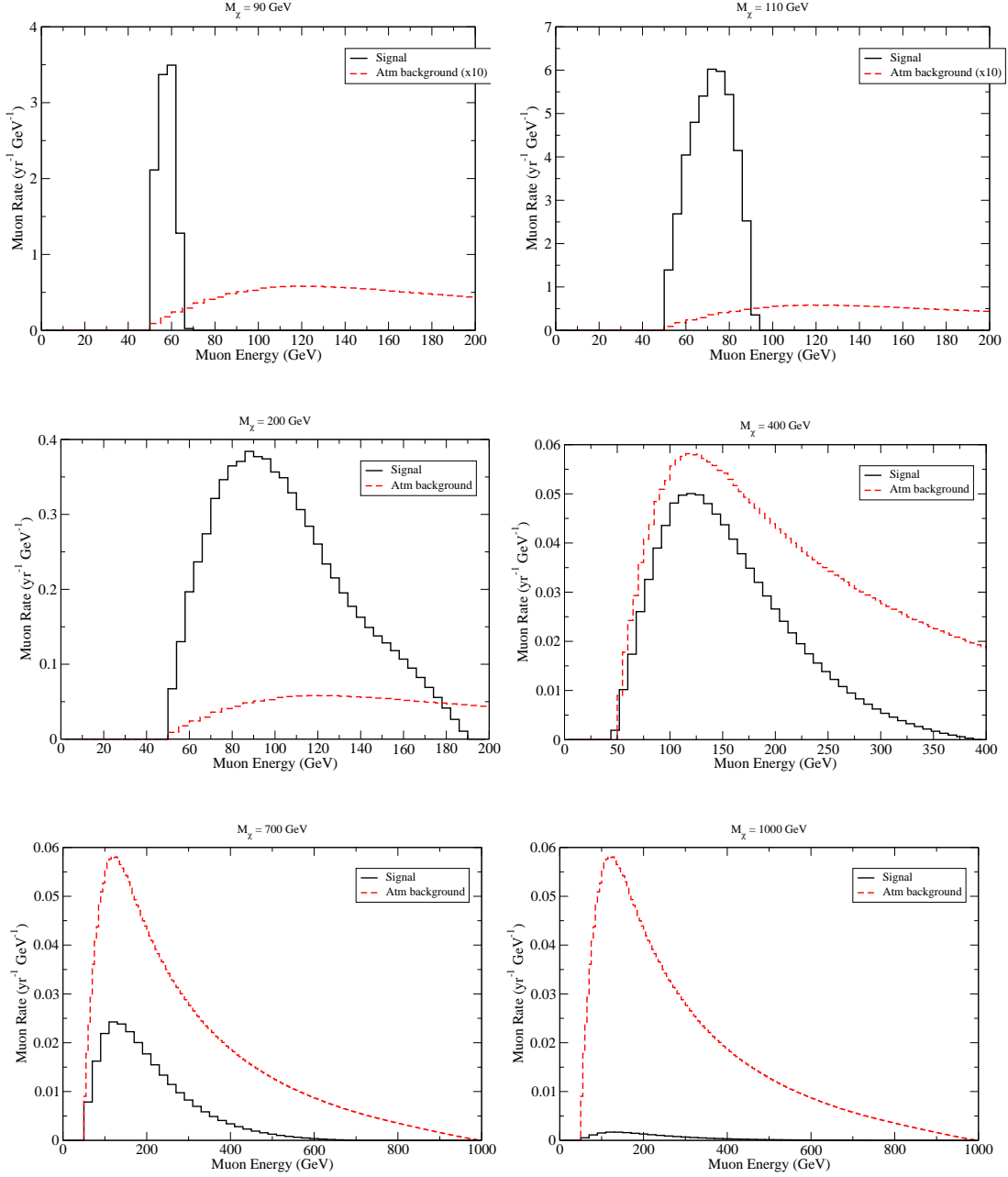

   \begin{center}
      \begin{tabular}{cc}
\vspace{0.2in}      \includegraphics[width=0.43\textwidth]{plots/pt01-muonrate.eps}\hspace{0.1in}
      \includegraphics[width=0.43\textwidth]{plots/pt02-muonrate.eps}\\\vspace{0.2in}
      \includegraphics[width=0.43\textwidth]{plots/pt03-muonrate.eps}
      \includegraphics[width=0.43\textwidth]{plots/pt04-muonrate.eps}\\
      \includegraphics[width=0.43\textwidth]{plots/pt06-muonrate.eps}
      \includegraphics[width=0.43\textwidth]{plots/pt05-muonrate.eps}
      \end{tabular}
      \caption{Muon event rates and atmospheric backgrounds in IceCube.}
      \label{fig:murate}
   \end{center}
\end{figure}

To severely reduce the background from atmospheric neutrinos,  we include only events within a narrow angular cone along the line of sight from the IceCube detector to the Sun.  We parameterize the zenith angle to of the Sun at the South Pole as
\be
\cos \theta_z = \cos \left[ {\pi\over2}+\theta_{dec}^{max} \sin \left(2\pi \tau\right)\right],
\ee
where $\theta_{dec}^{max}=23^\circ 26'$ is the maximum declination of the Sun in the celestial sphere throughout the year.  Here we have denoted the time of year by the dimensionless parameter $\tau$ which spans a full year with the values $0< \tau \le1$, where $\tau=0,\half$ correspond to the September and March Equinoxes, respectively.  The total muon rate along the line of sight to the Sun is given by
\bea
{d N_\mu\over dE_\mu} &=& \int_{E_\mu}^{\infty}  dE_{\nu_\mu}  \int_0^\half d\tau \left[{d \sigma^p_\nu (E_{\nu_\mu}, E_\mu)\over dE_\mu} \rho_{p}+{d \sigma^n_\nu (E_{\nu_\mu}, E_\mu)\over dE_\mu} \rho_{n}\right] \nn \\
&\times& R_\mu(E_\mu) A_{eff}\left(E_\mu\right) {d\Phi_\nu \over dE_{\nu_{\mu}}d\cos\theta_z }  {d\cos\theta_z\over d\tau} R(\cos\theta_z) \\\nn
&+& \left(\nu \to \bar \nu\right),
\eea
where $R(\cos\theta_z) = 0.70 - 0.48 \cos\theta_z$ is a detector efficiency for up-going muons that takes into account the rock bed below the detector and other angular dependence factors in IceCube~\cite{GonzalezGarcia:2005xw}.  The factor of $\half$ in the upper limit of the $\tau$ integration takes into account the time that the Sun is below the horizon.  The $1\sigma$ resolution of IceCube is $1^\circ$ ~\cite{Halzen:2007xx}; we therefore include the flux observed along the line of sight to the Sun within a cone of angular diameter $3^\circ$.  This dramatically reduces the  background from atmospheric neutrinos to that shown in Fig.~\ref{fig:murate}.

We show the signal muon event rate and atmospheric background~\cite{Honda:2006qj} in IceCube for six parameter points in Fig.~\ref{fig:murate} and Table \ref{tab:atmbkg}.  Defining the signal region of $E_{\mu}^{thr} < E_\mu < 300$ GeV, we determine the statistical significance of the six illustrative points in Table \ref{tab:atmbkg}\footnote{Note that smearing of the muon energy can yield a tail on the lower end of the spectrum. }.   This choice of the upper cut at 300 GeV is because of the large attenuation of high energy neutrinos by the CC and NC interactions in the Sun makes the signal contribution at higher energies small.  Here, we purposely chose the window to be independent of the neutralino mass, but once a signal is seen the window can be narrowed to improve the signal to background ratio.  In the above window with a few years of data, IceCube can make a $5 \sigma$ $\N_1$ discovery from $M_W$ up to 400 GeV.  The case for ${\cal O}(M_Z)$ are even better, easily yielding a discovery signal in the first year of data.

\begin{table}[htdp]
\caption{Muon rates in one year of IceCube data from neutralino annihilations and atmospheric neutrino backgrouns and the total statistical significance $\sigma_{stat}=S/\sqrt{B}$.  The event rates include all events where $E_{\mu}^{thr} < E_{\mu} < 300$ GeV, with $E_\mu^{thr} = 50$ GeV.}
\begin{center}
\begin{tabular}{|c|ccc|}
\hline
$M_{\N_1}$ (GeV) & $N_{\mu}^{Signal}$ & $N_{\mu}^{Bkg}$ & $\sigma_{stat}$ \\
\hline
90      &       41   & 10.4  & 13\\
110     &       170   & 10.4 & 53\\
200     &       29   & 10.4& 9.0\\
400     &       7.0    & 10.4 &2.2\\
700     &       4.1    & 10.4 & 1.3\\
1000    &       0.29    & 10.4 & 0.09\\
\hline
\end{tabular}
\end{center}
\label{tab:atmbkg}
\end{table}

From the signal in Fig.~\ref{fig:murate}, one can use the signal rate and the shape of the differential muon rate to extract further information on the nature of the neutralino .  A determination of the neutralino mass can be made from the shape of the muon energy distribution and the capture rate of neutralinos in the Sun can be determined from the total signal rate.  The capture rate in turn gives information on the local galatic DM density and its velocity (cf. Eq. (\ref{eq:caprate})) that can complement inferences from direct detection experiments.

\subsection{Limit Spectra}

When the DM particle is very massive and thus produces very energetic neutrinos, the features of the neutrino energy distribution tend to be washed out, due to the neutrinos interacting more with the solar matter.  This gives rise to the so-called ``limit spectra," which becomes a fair description for $E_\nu$ greater than 200 GeV or so \cite{Cirelli:2005gh}.  The limit spectra assumes that (i) oscillations and interactions with matter populate the different neutrino flavors uniformly so that the density matrix can be reduced to a scalar, (ii) the re-injection spectra from scattering is flat in $E'_\nu/E_\nu$, and (iii) the cross section is proportional to the energy of the incoming neutrino. In our case, the first  assumption is satisfied to zeroth order since the initial neutrino populations uniformly populate the three flavors.  However, $\tau$ regeneration breaks this uniformity and makes the $\nu_\tau$ population different than the $\nu_e$ population\footnote{Oscillations between the $\nu_\tau$ and $\nu_\mu$ neutrinos tend to keep these two flavors equally populated.}.  The reinjection spectra are preferentially skewed toward lower energies as calculated in Appendix \ref{apx:regen} and shown in Fig.~\ref{fig:transfn} and are not well approximated by flat spectra.  The third assumption is valid in the appropriate neutrino energy range of few GeV $\le E_\nu \le \text{few TeV} $.

\begin{figure}
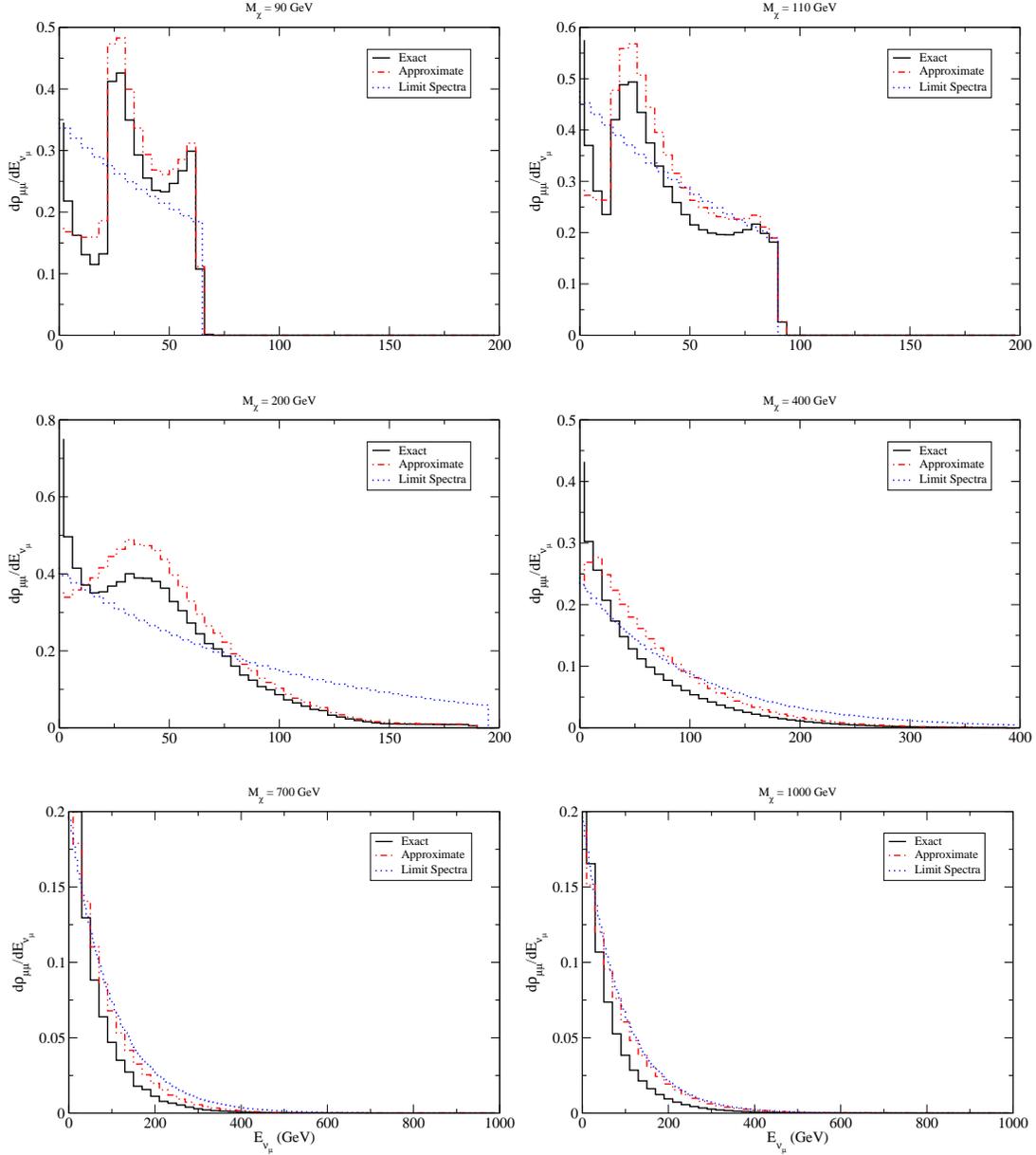

   \begin{center}
      \begin{tabular}{cc}
      \includegraphics[width=0.43\textwidth]{plots/pt01-limitspect.eps}\hspace{1mm}
      \includegraphics[width=0.43\textwidth]{plots/pt02-limitspect.eps}\\
      \includegraphics[width=0.43\textwidth]{plots/pt03-limitspect.eps}\hspace{1mm}
      \includegraphics[width=0.43\textwidth]{plots/pt04-limitspect.eps}\\
      \includegraphics[width=0.43\textwidth]{plots/pt06-limitspect.eps}
      \includegraphics[width=0.43\textwidth]{plots/pt05-limitspect.eps}
      \end{tabular}
      \caption{Given are our exact calculation for neutrinos (solid black line), the approximate spectra as described in~\cite{Cirelli:2005gh}, and the limit spectra.}
      \label{fig:limitspec}
   \end{center}
\end{figure}

For all $\nu$, the limit spectra can be determined analytically~\cite{Cirelli:2005gh}
\be
\label{eq:limitspect}
\rho(r,E) = e^{-{E \over {\cal E}}{r\over R_{\odot}}} \left\{\rho(r=0,E) + {1\over {\cal E}}\int_E^\infty \rho(r=0,E')dE'\right\}
\ee 
and approximated as a decaying exponential in the high energy limit
\be
\rho \left(r,E\right) \propto e^{-{E\over{\cal E}}{r\over R_{\odot}}}.
\ee
Here ${\cal E}$ is $\approx$ 100 GeV for neutrinos and $\approx$ 140 GeV for anti-neutrinos~\cite{Cirelli:2005gh}.  For very high neutrino energies, the spectral distributions become less distinctive and bear little resemblance to the injected spectra.  This effect unfortunately limits the ability to detect neutrinos of energies greater than $\approx$ 400 GeV or so, as can be seen in Fig.\ref{fig:limitspec}.  Combined with the afforementioned low-energy cutoff from the effective area of the detector, we are confined to a ``window" of possible muon detection energies from roughly 50 GeV to 400 GeV for IceCube.

\section{Conclusions}
\label{sect:conc}
We have made a comprehensive analysis of the prospects for observation in km$^2$ area detectors of high energy neutrinos from the annihilations of neutralino dark matter in the Sun.  The underlying assumptions of this study are the following:
\ben
\item The lightest stable neutralino of the mSUGRA model accounts for the dark matter density determined by the WMAP experiment
\item The neutralinos are gravitationally captured by the Sun and settle to the core of the Sun where they annihilate with the annihilation and capture rates in equilibrium.  We have assumed the standard estimate of the neutralino capture rate of Eq. (\ref{eq:caprate}) in estimating the signal.
\item The Focus Point/Hyperbolic Branch is the relevant region of the parameter space of the mSUGRA model,
since the high sfermion masses at the FP solves the SUSY flavor and CP problems and gives low fine tuning.
\item The mass of the lightest neutralino has mass above that of the W-boson, so that $WW, ZZ$, and $t\bar t$ are the dominant annihilation channels, giving high energy neutrinos from their decays.
\een

The Focus Point region gives large fluxes of high energy neutrinos for two reasons:
\ben
\item The spin-dependent cross-section for neutralino capture in the Sun is enhanced by a factor of $10^3$ over the spin-independent cross-section relevant to most current searches for direct detection of neutralinos through nuclear recoils.
\item The lightest neutralino is a bino-higgsino admixture which gives a large cross-section for the $WW$ and $ZZ$ production processes.
\item The contributions to the neutrino flux are dominated by the top quark contributions once the LSP mass exceeds the top quark mass.
\een

Our calculations included all the physically relevant processes in neutrino propagation through the Sun with the quantum mechanical density evolution Eq (\ref{eq:prop}), including:
\ben
\item Production at the center of the Sun with equal population of the three neutrino flavors.
\item Oscillations of three neutrinos with vacuum oscillations due to the atmospheric and solar mass-squared differences in the tribimaximal approximation to the mixing matrix which has a zero $\theta_{13}$ mixing angle and with MSW effects for non-zero $\theta_{13}$; MSW effects are not expected to be significant for high neutrino energy threshold of IceCube.  
\item Source terms due to the absorption and re-injection of neutrinos propagating through matter, including both CC and NC interactions that re-inject neutrinos with lower energy.  In particular, we include all neutrino decay products from the production of tau-leptons with exact calculations of the tau decay distributions.  We find that the tau-regeneration effects are not large, but that the absorption effects strongly diminish the event rates from neutralinos of mass above a few hundred GeV.
\een

Our study made major improvements over previous analyses:
\ben
\item We use take into account the full spin-dependence of the production and decay processes, which have important effects on the neutrino energy distributions.
\item We make quantitative predictions for the absolute neutrino rates, since we specialize to the specific FP region.
\item We include off-shell $W, Z$, and top decays in addition to on-shell.
\item We provide analytic formulas for on-shell decays and show that the numerical calculations of exact matrix elements obtained with the SMADGRAPH program are in agreement.
\een

We have compared our results with other analyses and approximations where possible:
\ben
\item We find that the approximation to the density evolution of Eq. (\ref{eq:limitspect})
is reasonably good for neutralino masses  of order 100 GeV, but shows substantial deviations from the exact results at higher neutralino masses:  see Fig. \ref{fig:limitspec}.
\item We find that the limit spectra from Cirelli et al Eq. (\ref{eq:limitspect}) is not so good in representing the exact results at any neutralino mass, even though it does qualitatively describe the fall-off with neutrino energy due to absorption effects.
\item The signal region is broadly defined in muon energy by the 50 GeV threshold cut and the rapid decrease at energies above 300 GeV from absorption effects.
\een

In the detection of muon-neutrino signals from the annihilations, we calculate event rates for the IceCube experiment, taking into account
\ben
\item A muon energy threshold of 50 GeV
\item The dependence of the detector area on muon energy due to the range of the muons in ice.
\item The backgrounds from atmospheric neutrinos as calculated from the flux in Ref. \cite{Honda:2006qj}; we include a 3 degree cone for the angular resolution along the line of sight from the Sun to the detector.
\een

The major conclusions of our study are as follows:

\ben
\item The signals from the annihilations of neutralinos in the mass range of 90 GeV to 400 GeV should be easily identifiable above the atmospheric neutrino backgrounds:  see Table \ref{tab:atmbkg}.
About 10 to 200 muons events are created from muon neutrinos by neutralino annihilations in the Sun on an atmospheric muon background of about 10 events (within a 3 degree angular cone the center of the Sun).  The breadth of the signal is in muon energies a measure of the neutralino mass.  The size of the signal is governed by the neutralino mass and its couplings to weak bosons and top-quark pairs as well as the capture rate of neutralinos in the Sun.

\item The IceCube experiment is most sensitive to neutralinos of mass around 110 GeV, above the thresholds for $WW$ and $ZZ$ production by the neutralino annihilation process.    However, the signals from neutralinos of mass $M_{\N_1} > m_t$, where $t\bar t$ annihilation dominates can also be large for neutralinos of mass $M_{\N_1}$ of order 200 GeV.

\item The absorption effects in the Sun likely preclude the detection in IceCube of neutralino annihilations for neutralino masses above 400 GeV.

\item The predicted large signals depend on several assumptions that could be overly optimistic: The Focus Point with large bino-higgsino mixing is the relevant region of mSUGRA parameter space (this is subject to validation at the LHC); The Spin-Dependent capture cross-section is much larger than the Spin-Independent Capture cross-section (this is subject to test in direct detection experiments of WIMP recoils); the estimated gravitational capture rate of neutralinos in the Sun is realistic (it could be enhanced by caustics or reduced if there is a dark matter density under-density in our galactic region).

\item The forthcoming data from IceCube and KM3 offer the prospect for discovery of neutralino DM if nature has cooperated.

\een

\begin{acknowledgments}
\end{acknowledgments}
We thank H. Baer, S. Desai, F. Halzen, P. Huber, C. Kao, and M. Maltoni for valuable discussions.  V.B. thanks the Aspen center for Physics for hospitality during the course of this work.  This work was supported in part by the U.S.~Department of Energy under grant Nos. DE-FG02-95ER40896 and DE-FG02-84ER40173, and by the Wisconsin Alumni Research Foundation.

\appendix

\section{Phase space partitioning}
\label{apx:ps}
The calculations of the cross sections $\sigma (\chi \chi \to W e \overline \nu)$ and $\sigma (\chi \chi \to Z \nu \overline \nu)$ are quite similar. Here, we compute the cross section for $WW^*$ production, and then give the corresponding expressions for the $ZZ^*$ process. The cross section $\sigma (\chi \chi \to W e \overline \nu)$ is given by
\be
\sigma = \frac{1}{2 \lambda^{\frac{1}{2}}(s,p^2_{\chi_{1}},p^2_{\chi_{2}})(2 \pi)^5} \sum |{\cal M}^2| d_s(PS \chi \chi \to W e \overline \nu),
\ee
where $\lambda$ is the triangular function
\be
\lambda (a,b,c) \equiv a^2 + b^2 + c^2 - 2(ab + ac + bc).
\ee
The phase space is 
\be
d_3(PS \chi \chi \to W e \overline \nu) = \left( \frac{\pi}{2} \right) ^2 \left( \frac{d \Omega}{4 \pi} \right) ^2 \lambda^{\frac{1}{2}}(1,\frac{m^2_X}{s},\frac{m^2_W}{s}) \lambda^{\frac{1}{2}}(1,\frac{m^2_e}{m^2_X},\frac{m^2_{\nu}}{m^2_X}) d m^2_X,
\ee
where we integrate over the range $m_e + m_{\nu} \leq m_X \leq \sqrt{s} - m_W.$ The corresponding calculation for $ZZ^*$, with $m_{\nu} = 0$, yields
\be
d_3(PS \chi \chi \to Z \nu \overline \nu) = \left( \frac{\pi}{2} \right) ^2 \left( \frac{d \Omega}{4 \pi} \right) ^2 \lambda^{\frac{1}{2}}(1,\frac{m^2_X}{s},\frac{m^2_Z}{s}) d m^2_X,
\ee
where $0 \leq m_X \leq \sqrt{s} - m_Z$.

\section{Neutralino number in the Solar Core}
\label{apx:neutdens}
It is interesting to estimate  the number of the accumulated  dark matter particles in the solar core.  The content of the dark matter located at the core by the equilibrium condition can be modeled by~\cite{Halzen:2005ar}
\be
N^2={C_\odot\over A_\odot}={{C^\odot V}  \over \langle \sigma v\rangle}
\ee
The effective volume $V$ of the DM core of the Sun is estimated to be~\cite{Griest:1986yu,Gould:1987ir}
\be
V=5.7 \times 10^{27} \hbox{ cm}^3 (100 \hbox{ GeV}/M_{\N_1})^{3\over2} 
\ee
The annihilation cross section is dimensionally estimated as 
\be
\langle \sigma v\rangle \approx B \alpha^2/M_{\N_1}^2 
\ee 
with the coefficient $B$ of order 1.  For $M_{\N_1}=100$ GeV, We find $N\approx 10^{38}$ for the SD capture rate. This number is  many orders of magnitude larger than the average number $\bar N$ of dark matter particles 
enclosed by a volume in the space of the size of the Sun, $\bar N\approx 4\times 10^{30}$.

\section{Neutrino-nucleon scattering}
\label{apx:neutscatt}
Neutrinos scatter off nucleons via the NC or CC interactions.  The cross section expressions are relatively simple for neutrinos  in the energy range we consider (${\cal O}(1\text{ GeV}) \le E_{\nu}\le {\cal O}(1\text{ TeV})$) where the $W$ and $Z$ propagator effects can be neglected since $Q^2 \ll M_W^2$.

\subsection{CC interaction}
For CC scattering, the neutrino-quark cross sections are given by \cite{Winter:2000xx,Barger:1987nn}
\bea
{d \hat \sigma\over d y}(\nu d \to \ell u) &=& {d \hat \sigma\over d y}(\bar \nu \bar d \to \bar\ell \bar u) = {G_F^2 \hat s \over \pi},\\
{d \hat \sigma\over d y}(\nu \bar d \to \ell \bar u) &=& {d \hat \sigma\over d y}(\bar \nu d \to \bar\ell u) = {G_F^2 \hat s \over \pi}(1-y)^2,
\eea
where $y=1-E_{\ell}/E_{\nu}$, $\sqrt {\hat s} = \sqrt{s x}$ is the CM energy of the subprocess and $x$ is the fraction of nucleon momenta imparted on the quark.  The resulting scattering off a proton is given by~\cite{Strumia:2006db}
\bea
{d \sigma\over d y}(\nu p \to \ell X) &=& \int_0^1 dx {d \hat \sigma\over d y}(\nu \bar u \to \ell \bar d) f_{\bar u/p}(x)+ {d \hat \sigma\over d y}(\nu d \to \ell u) f_{d/p}(x) \\\nn
&\simeq& {2 m_p E_\nu G_F^2  \over \pi}(0.15 + 0.04 (1-y)^2),\\
{d \sigma\over d y}(\bar \nu p \to \bar \ell X) &=& \int_0^1 dx {d \hat \sigma\over d y}(\bar \nu \bar d \to  \bar \ell \bar u) f_{\bar d/p}(x)+ {d \hat \sigma\over d y}(\bar\nu u \to \bar\ell d) f_{u/p}(x)\\\nn
& \simeq& {2 m_p E_\nu G_F^2  \over \pi}(0.06 + 0.25 (1-y)^2),
\eea
where $f_{q/N}(x)$ is the parton density function for parton $q$ in nucleon $N$.  The corresponding neutron differential cross sections are~\cite{Strumia:2006db}
\bea
{d \sigma\over d y}(\nu n \to \ell X) &=& \int_0^1 dx {d \hat \sigma\over d y} (\nu \bar u \to \ell \bar d)f_{\bar u/n}(x)+ {d \hat \sigma\over d y}(\nu d \to \ell u) f_{d/n}(x) \\\nn
& \simeq& {2 m_n E_\nu G_F^2  \over \pi}(0.25 + 0.06 (1-y)^2),\\
{d \sigma\over d y}(\bar \nu n \to \bar \ell X) &=& \int_0^1 dx {d \hat \sigma\over d y}(\bar \nu \bar d \to  \bar \ell \bar u) f_{\bar d/n}(x)+ {d \hat \sigma\over d y}(\bar\nu u \to \bar\ell d) f_{u/n}(x) \\\nn
& \simeq& {2 m_n E_\nu G_F^2  \over \pi}(0.04 + 0.15 (1-y)^2).
\eea

\subsection{NC interaction}
The NC scattering is calculated in the same way.  The subprocess are given by
\bea
{d \hat \sigma\over d y}(\nu q \to \nu' q') &=& {d \hat \sigma\over d y}(\bar \nu \bar q \to \bar\nu' \bar q') = {G_F^2 \hat s \over \pi}\left(g_{Lq}^2+g_{Rq}^2(1-y)^2\right),\\
{d \hat \sigma\over d y}(\bar \nu q \to \bar \nu q') &=& {d \hat \sigma\over d y}(\nu \bar q \to\nu' \bar q') = {G_F^2 \hat s \over \pi}\left(g_{Rq}^2+g_{Lq}^2(1-y)^2\right),
\eea
where $y=1-E_{\nu'}/E_{\nu}$ and the $Z$ boson couplings are
\be
g_{Lu} = \half -{2\over3}\sin^2\theta_W,\qquad g_{Ru} = -{2\over 3}\sin^2\theta_W,\qquad g_{Ld} = -\half +{1\over3}\sin^2\theta_W,\qquad g_{Rd} = {1\over 3}\sin^2\theta_W.
\ee
The total differential cross sections for proton target are then given by~\cite{Strumia:2006db}
\bea
{d \sigma\over d y}(\nu p \to \nu' X) &=& \int_0^1 dx \sum_{q=u,d,\bar u,\bar d}{d \hat \sigma\over d y} (\nu q \to \nu' q')f_{q/p}(x)\\\nn
& \simeq& {2 m_p E_\nu G_F^2  \over \pi}\left[0.058+0.022(1-y)^2\right],\\
{d \sigma\over d y}(\bar \nu p \to \bar \nu' X) &=& \int_0^1 dx \sum_{q=u,d,\bar u,\bar d}{d \hat \sigma\over d y} (\bar \nu q \to \bar\nu' q')f_{q/p}(x)\\\nn
& \simeq& {2 m_p E_\nu G_F^2  \over \pi}\left[0.022+0.058(1-y)^2\right],
\eea
and for a neutron target by
\bea
{d \sigma\over d y}(\nu n \to \nu' X) &=& \int_0^1 dx \sum_{q=u,d,\bar u,\bar d}{d \hat \sigma\over d y} (\nu q \to \nu' q')f_{q/n}(x)\\\nn
& \simeq& {2 m_n E_\nu G_F^2  \over \pi}\left[0.064+0.019(1-y)^2\right],\\
{d \sigma\over d y}(\bar \nu n \to \bar \nu' X) &=& \int_0^1 dx \sum_{q=u,d,\bar u,\bar d}{d \hat \sigma\over d y} (\bar \nu q \to \bar\nu' q')f_{q/n}(x)\\\nn
& \simeq& {2 m_n E_\nu G_F^2  \over \pi}\left[0.019+0.064(1-y)^2\right].
\eea

\section{$\tau$ regeneration}
\label{apx:regen}
To account for $\tau$ regeneration, we fold the CC production cross section of leptons with  decay distribution of the $\tau$
\bea 
f_{\nu_\tau \to \nu_\tau}(u) &=& N \int_u^1 \left(1+{z^2 \over 5}\right)  \sum_i BF_i\left(g_{0i}\left({u \over z}\right) + P g_{1i}\left({u \over z}\right)\right){1 \over z}dz,\\
f_{\bar \nu_\tau \to \bar \nu_\tau}(u) &=&N \int_u^1 \left({1 \over 5}+z^2\right)  \sum_i BF_i\left(g_{0i}\left({u \over z}\right) + P g_{1i}\left({u \over z}\right)\right){1 \over z}dz,
\label{eqn:transfer}
\eea
with similar expressions for $f_{\bar \nu_\tau \to \nu_{e,\mu}}$ and $f_{\nu_\tau \to \bar \nu_{e,\mu}}$.  Here, the factors of 1/5~\cite{Barger:1999fs} approximate the more precise values in Appendix \ref{apx:neutscatt}.  In this parameterization, $u = E_\nu^{out}/E_\nu^{in}$ and $z = E_\tau/E_\nu^{in}$.  The normalization factor $N$ is chosen such that 
\bea
N &=& \int_0^1 du f_{\nu \to \nu} (u) = \left\{\begin{array}{cc}
1	& \nu_\tau \to \nu_\tau\\
0.18& \nu_\tau \to \bar \nu_{e,\mu}
\end{array}\right.
\eea
The first factor in each equation arises from the CC production cross section for $\tau^-$ and $\tau^+$, respectively. The sum is over each contributing mode (note that for modes with $\nu_e$ or $\nu_\mu$ in the final state, there is only one term in the sum), and $BF_i$ is the corresponding branching fraction.  $P$ is the polarization of the decaying $\tau$, and is $\pm 1$ for $\tau^\mp$.  $g(y)$ is the energy spectra of the neutrinos from the decaying $\tau$.  We now calculate this spectra for the modes $\nu_\tau \ell \bar \nu_\ell$, $\nu_\tau e/\mu \bar \nu_{e/\mu}$ and $\nu_\tau \pi$, and give a table of all results.  

To calculate the spectra, let $f_0(x)$ be the $\nu_\tau$ distribution in the static frame of the parent $\tau$, (e.g. 
$f_0(x)=2x^2(3-2x)$ for $\nu_\tau$ from $\tau$).  The distribution is transformed into $g(y)$ in the fragmentation
frame, with $y=(x/2) (1+\beta \cos\theta)$, where $\beta$ is the velocity and $\theta$ is the polar angle.  Each $x$-bin at $x_0$ or $\delta(x-x_0)$ contributes a flat $y$
distribution  $\theta(x_0-y)/x_0$ in the limit $\beta\to 1$ for an unpolarized parent.  Therefore,
\be 
g_0(y)=\int_0^1 dx  f_0(x) \theta(x-y) /x=\int_y^1 dx f_0(x)/x \ , 
\ee
For $\nu_\tau $ from $\tau$, we have
\be
g_0(y)=\int_y^1 2x(3-2x)dx = {5\over3}-3y^2+{4\over3}y^3
\ee
Another way to obtain the result is
\be 
g_0(y)=\int_{-1}^1 \hbox{$1\over2$} d\cos\phi \int_0^1  dx f_0(x) \delta (y-\hbox{$1\over2$} x(1+\cos\phi)) \ .
\ee

If the polarization effect in the rest frame is defined by 
\be
{1\over N}   {dN \over dx d\cos\theta} = f_0(x)+f_1(x)\cos\theta
\ee 
which, in the laboratory frame, can be parameterized by~\cite{Dutta:2000hh,Pasquali:1998xf}
\be{1\over N}   {dN \over dy } =g_0(y)+P g_1(y),
\ee
then
\be
g_1(y)=+\int_y^1 dx (2y-x)f_1(x)/x^2
\ee
For the $\nu_\tau$ from $\tau \to \nu_\tau + \mu +\bar\nu_\mu $,
\be
f_1(x)=-2x^2(2x-1) 
\ee
and
\be
g_1(y)=\int_y^1 2(1-2x)(2y-x) dx = {1\over3}+{8\over3}y^3-3y^2
\ee 

We now move to the case of  $\bar\nu_\mu $ in the $\tau$ decay.
The transition probability is \cite{Bjorken:1979dk}
\be
\sum|{\cal M}|^2=
 64 G_F^2 \  (\mu\cdot \nu_\tau)\ \ \bar \nu_\mu\cdot (\tau-m_\tau S_\tau)
=16 G_F^2 m_\tau^4 (1-x)x(1+\cos\theta)
\ee
\be
f_0(x)=12 x^2(1-x) \ ,\quad  f_1(x)=+12 x^2(1-x)
\ee
Then 
\bea
g_0(y)&=&\int_y^1 12(1-x)x dx=2-6y^2+4y^3\\
g_1(y)&=&\int_y^1 dx 12 (1-x)(2y-x)= -2+12y-18y^2+8y^3
\eea
in agreement with Eq.(96) in ~\cite{Lipari:1993hd}.
Note that the corresponding spectra for the $\tau^+$ decay are
obtained with the substitution $g_1^{\tau^+}(y) =-g_1^{\tau^-}(y)$.

\begin{table}[htdp]
\caption{Fragmentation functions for various decay modes of the $\tau^-$ lepton, with $y={E_{\nu_\tau}\over E_{\tau}}$ and $r_X = m_X^2/m_\tau^2$.}
\begin{center}
\begin{tabular}{|c|c|c|c|}
\hline
$\tau^-$ decay mode & BF &$g_0(y)$ & $g_1(y)$\\
\hline
$\nu_\tau \ell \bar \nu_\ell$  & 0.18 &${5\over3}-3y^2+{4\over3}y^3 $&$ {1\over3}+{8\over3}y^3-3y^2$\\
$\nu_\tau \pi$ & 0.12 & ${1\over 1-r_\pi}\theta(1-r_\pi-y)$&$ -{2y-1+r_\pi \over (1-r_\pi)^2}\theta(1-r_\pi-y)$\\
$\nu_\tau a_1$ & 0.13 & ${1\over 1-r_{a_1}}\theta(1-r_{a_1}-y)$& $ -{2y-1+r_{a_1} \over1-r_{a_1}}{1-2r_{a_1}\over1+2r_{a_1}}\theta(1-r_{a_1}-y)$\\
$\nu_\tau \rho$ & 0.26 & ${1\over 1-r_\rho}\theta(1-r_\rho-y)$& $ -{2y-1+r_\rho \over 1-r_\rho}{1-2r_\rho\over1+2r_\rho}\theta(1-r_\rho-y)$\\
$\nu_\tau X$ & 0.13 & ${1\over 0.3}\theta(0.3-y)$& 0 \\
\hline
\end{tabular}
\end{center}
\label{default}
\end{table}%

\begin{table}[htdp]
\caption{Fragmentation functions for various decay modes of the $\tau^-$ lepton, with $y={E_{\bar \nu_{e,\mu}}\over E_{\tau}}$ and $r_X = m_X^2/m_\tau^2$ \cite{Dutta:2000jv}.  }
\begin{center}
\begin{tabular}{|c|c|c|c|}
\hline
$\tau^-$ decay mode & BF &$g_0(y)$ & $g_1(y)$\\
\hline
$\nu_\tau \ell \bar \nu_\ell$  & 0.18 & $2-6y^2+4y^3$ &$-2+12y-18y^2+8y^3$ \\
\hline
\end{tabular}
\end{center}
\label{default}
\end{table}%

For the mode $\tau \to \pi^-\nu_\tau$, we define
$$  E_{\nu_\tau}={m_\tau^2-m_\pi^2\over 2 m_\tau} \ ,\quad
x={2E_{\nu_\tau} \over m_\tau}            \ ,\quad
r_\pi={m_\pi^2\over m_\tau^2} \ ,\quad
x=1-r_\pi 
$$
The rest frame distribution is a delta function,
$f_0(x)=\delta(x-(1-r_\pi))$.
We can follow what we did before to obtain the boost fragmentation
function,
\be
g_0(y)=\int_y^1 dx f_0(x)/x= \int_y^1 dx \delta(x-(1-r_\pi))/x 
=  \theta(1-r_\pi-y)/(1-r_\pi)
\ee
This agrees with formulas by Ref. \cite{Dutta:2000jv}.  For the polarized distribution, the angular distribution in the parent rest frame is $\propto 1-\cos\theta$, so,
\be
g_1(y)=-\int_y^1 dx(2y-x)\delta(x-1+r_\pi)/x^2=-(2y-1+r_\pi)\theta(1-r_\pi-y)/(1-r_\pi^2).
\ee

\bibliographystyle{h-physrev}
\bibliography{nusignals}                      

\begin{thebibliography}{100}

\bibitem{Kolb:1990vq}
E.~W. Kolb and M.~S. Turner,
\newblock Front. Phys. {\bf 69}, 1 (1990), The Early universe.

\bibitem{Drees:2004jm}
M.~Drees, R.~Godbole, and P.~Roy,
\newblock Theory and phenomenology of sparticles: An account of four-
  dimensional N=1 supersymmetry in high energy physics,
\newblock Hackensack, USA: World Scientific (2004) 555 p.

\bibitem{Baer:2006rs}
H.~Baer and X.~Tata,
\newblock Weak scale supersymmetry: From superfields to scattering events,
\newblock Cambridge, UK: Univ. Pr. (2006) 537 p.

\bibitem{Binetruy:2006ad}
P.~Binetruy,
\newblock Supersymmetry: Theory, experiment and cosmology,
\newblock Oxford, UK: Oxford Univ. Pr. (2006) 520 p.

\bibitem{Haber:1984rc}
H.~E. Haber and G.~L. Kane,
\newblock Phys. Rept. {\bf 117}, 75 (1985).

\bibitem{Chamseddine:1982jx}
A.~H. Chamseddine, R.~Arnowitt, and P.~Nath,
\newblock Phys. Rev. Lett. {\bf 49}, 970 (1982).

\bibitem{Barger:1992ac}
V.~D. Barger, M.~S. Berger, and P.~Ohmann,
\newblock Phys. Rev. {\bf D47}, 1093 (1993), hep-ph/9209232.

\bibitem{Kane:1993td}
G.~L. Kane, C.~F. Kolda, L.~Roszkowski, and J.~D. Wells,
\newblock Phys. Rev. {\bf D49}, 6173 (1994), hep-ph/9312272.

\bibitem{Barger:1993gh}
V.~D. Barger, M.~S. Berger, and P.~Ohmann,
\newblock Phys. Rev. {\bf D49}, 4908 (1994), hep-ph/9311269.

\bibitem{Ellis:2003cw}
J.~R. Ellis, K.~A. Olive, Y.~Santoso, and V.~C. Spanos,
\newblock Phys. Lett. {\bf B565}, 176 (2003), hep-ph/0303043.

\bibitem{Abel:2000vs}
S.~Abel {\em et~al.},
\newblock (2000), hep-ph/0003154.

\bibitem{DMSAG}
Dark Matter Scientific Assessment Group ,
  http://www.science.doe.gov/hep/DMSAGReportJuly18,2007.pdf.

\bibitem{Jungman:1995df}
G.~Jungman, M.~Kamionkowski, and K.~Griest,
\newblock Phys. Rept. {\bf 267}, 195 (1996), hep-ph/9506380.

\bibitem{Strumia:2006db}
A.~Strumia and F.~Vissani,
\newblock hep-ph/0606054.

\bibitem{Mena:2007ty}
O.~Mena, S.~Palomares-Ruiz, and S.~Pascoli,
\newblock (2007), arxiv:0706.3909.

\bibitem{Baltz:2006fm}
E.~A. Baltz, M.~Battaglia, M.~E. Peskin, and T.~Wizansky,
\newblock Phys. Rev. {\bf D74}, 103521 (2006), hep-ph/0602187.

\bibitem{Hooper:2003ka}
D.~Hooper and L.-T. Wang,
\newblock Phys. Rev. {\bf D69}, 035001 (2004), hep-ph/0309036.

\bibitem{Arnowitt:2006jq}
R.~Arnowitt, B.~Dutta, T.~Kamon, N.~Kolev, and D.~Toback,
\newblock Phys. Lett. {\bf B639}, 46 (2006), hep-ph/0603128.

\bibitem{Roszkowski:2007va}
L.~Roszkowski, R.~R. de~Austri, J.~Silk, and R.~Trotta,
\newblock (2007), arxiv:0707.0622.

\bibitem{Jungman:1994jr}
G.~Jungman and M.~Kamionkowski,
\newblock Phys. Rev. {\bf D51}, 328 (1995), hep-ph/9407351.

\bibitem{Barger:2001ur}
V.~D. Barger, F.~Halzen, D.~Hooper, and C.~Kao,
\newblock Phys. Rev. {\bf D65}, 075022 (2002), hep-ph/0105182.

\bibitem{Halzen:2005ar}
F.~Halzen and D.~Hooper,
\newblock Phys. Rev. {\bf D73}, 123507 (2006), hep-ph/0510048.

\bibitem{GonzalezGarcia:2005xw}
M.~C. Gonzalez-Garcia, F.~Halzen, and M.~Maltoni,
\newblock Phys. Rev. {\bf D71}, 093010 (2005), hep-ph/0502223.

\bibitem{icecube:2001aa}
J.~Ahrens {\em et~al.},
\newblock (2001), IceCube Preliminary Design Document.

\bibitem{Sapienza:2005tz}
P.~Sapienza,
\newblock Nucl. Phys. Proc. Suppl. {\bf 145}, 331 (2005).

\bibitem{Aguilar:2006rm}
ANTARES, J.~A. Aguilar {\em et~al.},
\newblock Astropart. Phys. {\bf 26}, 314 (2006), astro-ph/0606229.

\bibitem{Resvanis:2006sb}
NESTOR, L.~K. Resvanis,
\newblock J. Phys. Conf. Ser. {\bf 39}, 447 (2006).

\bibitem{nestor:2007xx}
Nestor Collab. , http://www.uoa.gr/~nestor/.

\bibitem{Antares:2007xx}
Antares Collab. , http://antares.in2p3.fr/.

\bibitem{KM3:2007xx}
http://www.km3net.org/.

\bibitem{Cirelli:2005gh}
M.~Cirelli {\em et~al.},
\newblock Nucl. Phys. {\bf B727}, 99 (2005), hep-ph/0506298.

\bibitem{Chan:1997bi}
K.~L. Chan, U.~Chattopadhyay, and P.~Nath,
\newblock Phys. Rev. {\bf D58}, 096004 (1998), hep-ph/9710473.

\bibitem{Feng:1999mn}
J.~L. Feng, K.~T. Matchev, and T.~Moroi,
\newblock Phys. Rev. Lett. {\bf 84}, 2322 (2000), hep-ph/9908309.

\bibitem{Gould:1992xx}
A.~Gould,
\newblock Astrophys. J. {\bf 388} (1992).

\bibitem{Sikivie:2001fg}
P.~Sikivie,
\newblock Phys. Lett. {\bf B567}, 1 (2003), astro-ph/0109296.

\bibitem{Bertin:2002ky}
V.~Bertin, E.~Nezri, and J.~Orloff,
\newblock Eur. Phys. J. {\bf C26}, 111 (2002), hep-ph/0204135.

\bibitem{PhysRevD.17.2369}
L.~Wolfenstein,
\newblock Phys. Rev. D {\bf 17}, 2369 (1978).

\bibitem{PhysRevD.22.2718}
V.~Barger, K.~Whisnant, S.~Pakvasa, and R.~J.~N. Phillips,
\newblock Phys. Rev. D {\bf 22}, 2718 (1980).

\bibitem{Mikheev:1986gs}
S.~P. Mikheev and A.~Y. Smirnov,
\newblock Sov. J. Nucl. Phys. {\bf 42}, 913 (1985).

\bibitem{Lehnert:2007fv}
R.~Lehnert and T.~J. Weiler,
\newblock (2007), arXiv:0708.1035 [hep-ph].

\bibitem{Yao:2006px}
W.~M. Yao {\em et~al.},
\newblock J. Phys. {\bf G33}, 1 (2006).

\bibitem{Crotty:2002mv}
P.~Crotty,
\newblock Phys. Rev. {\bf D66}, 063504 (2002), hep-ph/0205116.

\bibitem{Spergel:2006hy}
WMAP, D.~N. Spergel {\em et~al.},
\newblock Astrophys. J. Suppl. {\bf 170}, 377 (2007), astro-ph/0603449.

\bibitem{Feng:1999zg}
J.~L. Feng, K.~T. Matchev, and T.~Moroi,
\newblock Phys. Rev. {\bf D61}, 075005 (2000), hep-ph/9909334.

\bibitem{Feng:2000bp}
J.~L. Feng and K.~T. Matchev,
\newblock Phys. Rev. {\bf D63}, 095003 (2001), hep-ph/0011356.

\bibitem{Baer:2005ky}
H.~Baer, T.~Krupovnickas, S.~Profumo, and P.~Ullio,
\newblock JHEP {\bf 10}, 020 (2005), hep-ph/0507282.

\bibitem{Alwall:2007st}
J.~Alwall {\em et~al.},
\newblock (2007), arXiv:0706.2334 [hep-ph].

\bibitem{Maltoni:2002qb}
F.~Maltoni and T.~Stelzer,
\newblock JHEP {\bf 02}, 027 (2003), hep-ph/0208156.

\bibitem{Cho:2006sx}
G.~C. Cho {\em et~al.},
\newblock Phys. Rev. {\bf D73}, 054002 (2006), hep-ph/0601063.

\bibitem{Stelzer:1994ta}
T.~Stelzer and W.~F. Long,
\newblock Comput. Phys. Commun. {\bf 81}, 357 (1994), hep-ph/9401258.

\bibitem{Murayama:1992gi}
H.~Murayama, I.~Watanabe, and K.~Hagiwara,
\newblock KEK-91-11.

\bibitem{Baer:1995nq}
H.~Baer, C.-h. Chen, F.~Paige, and X.~Tata,
\newblock Phys. Rev. {\bf D52}, 2746 (1995), hep-ph/9503271.

\bibitem{Baer:1995va}
H.~Baer, C.-h. Chen, F.~Paige, and X.~Tata,
\newblock Phys. Rev. {\bf D53}, 6241 (1996), hep-ph/9512383.

\bibitem{Baer:2007ya}
H.~Baer, V.~Barger, G.~Shaughnessy, H.~Summy, and L.-t. Wang,
\newblock (2007), hep-ph/0703289.

\bibitem{Ellis:1998kh}
J.~R. Ellis, T.~Falk, and K.~A. Olive,
\newblock Phys. Lett. {\bf B444}, 367 (1998), hep-ph/9810360.

\bibitem{Gomez:1999dk}
M.~E. Gomez, G.~Lazarides, and C.~Pallis,
\newblock Phys. Rev. {\bf D61}, 123512 (2000), hep-ph/9907261.

\bibitem{Lahanas:1999uy}
A.~B. Lahanas, D.~V. Nanopoulos, and V.~C. Spanos,
\newblock Phys. Rev. {\bf D62}, 023515 (2000), hep-ph/9909497.

\bibitem{Baer:2002fv}
H.~Baer, C.~Balazs, and A.~Belyaev,
\newblock JHEP {\bf 03}, 042 (2002), hep-ph/0202076.

\bibitem{Baer:1995nc}
H.~Baer and M.~Brhlik,
\newblock Phys. Rev. {\bf D53}, 597 (1996), hep-ph/9508321.

\bibitem{Barger:1997kb}
V.~D. Barger and C.~Kao,
\newblock Phys. Rev. {\bf D57}, 3131 (1998), hep-ph/9704403.

\bibitem{Baer:2003yh}
H.~Baer and C.~Balazs,
\newblock JCAP {\bf 0305}, 006 (2003), hep-ph/0303114.

\bibitem{Drees:1992am}
M.~Drees and M.~M. Nojiri,
\newblock Phys. Rev. {\bf D47}, 376 (1993), hep-ph/9207234.

\bibitem{Baer:1997ai}
H.~Baer and M.~Brhlik,
\newblock Phys. Rev. {\bf D57}, 567 (1998), hep-ph/9706509.

\bibitem{Baer:2000jj}
H.~Baer {\em et~al.},
\newblock Phys. Rev. {\bf D63}, 015007 (2001), hep-ph/0005027.

\bibitem{Ellis:2001ms}
J.~R. Ellis, T.~Falk, G.~Ganis, K.~A. Olive, and M.~Srednicki,
\newblock Phys. Lett. {\bf B510}, 236 (2001), hep-ph/0102098.

\bibitem{Roszkowski:2001sb}
L.~Roszkowski, R.~Ruiz~de Austri, and T.~Nihei,
\newblock JHEP {\bf 08}, 024 (2001), hep-ph/0106334.

\bibitem{Djouadi:2001yk}
A.~Djouadi, M.~Drees, and J.~L. Kneur,
\newblock JHEP {\bf 08}, 055 (2001), hep-ph/0107316.

\bibitem{Lahanas:2001yr}
A.~B. Lahanas and V.~C. Spanos,
\newblock Eur. Phys. J. {\bf C23}, 185 (2002), hep-ph/0106345.

\bibitem{Barger:1993vu}
V.~D. Barger, M.~S. Berger, P.~Ohmann, and R.~J.~N. Phillips,
\newblock Phys. Lett. {\bf B314}, 351 (1993), hep-ph/9304295.

\bibitem{Bardeen:1993rv}
W.~A. Bardeen, M.~S. Carena, S.~Pokorski, and C.~E.~M. Wagner,
\newblock Phys. Lett. {\bf B320}, 110 (1994), hep-ph/9309293.

\bibitem{Moroi:2006fp}
T.~Moroi, Y.~Sumino, and A.~Yotsuyanagi,
\newblock Phys. Rev. {\bf D74}, 015016 (2006), hep-ph/0605181.

\bibitem{Barger:2006gw}
V.~Barger, W.-Y. Keung, H.~E. Logan, and G.~Shaughnessy,
\newblock Phys. Rev. {\bf D74}, 075005 (2006), hep-ph/0608215.

\bibitem{Barger:1987nn}
V.~D. Barger and R.~J.~N. Phillips,
\newblock COLLIDER PHYSICS,
\newblock REDWOOD CITY, USA: ADDISON-WESLEY (1987) 592 P. (FRONTIERS IN
  PHYSICS, 71).

\bibitem{Chang:1992tu}
D.~Chang and W.-Y. Keung,
\newblock Phys. Lett. {\bf B305}, 261 (1993), hep-ph/9301265.

\bibitem{Gunion:1989we}
J.~F. Gunion, H.~E. Haber, G.~L. Kane, and S.~Dawson,
\newblock THE HIGGS HUNTER'S GUIDE,
\newblock ADDISON-WESLEY (1989) 425 P. (FRONTIERS IN PHYSICS, 80).

\bibitem{Barger:2005ve}
V.~Barger, W.-Y. Keung, H.~E. Logan, G.~Shaughnessy, and A.~Tregre,
\newblock Phys. Lett. {\bf B633}, 98 (2006), hep-ph/0510257.

\bibitem{Labonne:2006hk}
B.~Labonne, E.~Nezri, and J.~Orloff,
\newblock Eur. Phys. J. {\bf C47}, 805 (2006), hep-ph/0602111.

\bibitem{Raffelt:1992uj}
G.~Raffelt, G.~Sigl, and L.~Stodolsky,
\newblock Phys. Rev. Lett. {\bf 70}, 2363 (1993), hep-ph/9209276.

\bibitem{Bahcall:2004yr}
J.~N. Bahcall, S.~Basu, M.~Pinsonneault, and A.~M. Serenelli,
\newblock Astrophys. J. {\bf 618}, 1049 (2005), astro-ph/0407060.

\bibitem{Pontecorvo:1967fh}
B.~Pontecorvo,
\newblock Sov. Phys. JETP {\bf 26}, 984 (1968).

\bibitem{Gribov:1968kq}
V.~N. Gribov and B.~Pontecorvo,
\newblock Phys. Lett. {\bf B28}, 493 (1969).

\bibitem{Harrison:2002er}
P.~F. Harrison, D.~H. Perkins, and W.~G. Scott,
\newblock Phys. Lett. {\bf B530}, 167 (2002), hep-ph/0202074.

\bibitem{Nihei:2002ij}
T.~Nihei, L.~Roszkowski, and R.~Ruiz~de Austri,
\newblock JHEP {\bf 03}, 031 (2002), hep-ph/0202009.

\bibitem{Angle:2007uj}
J.~Angle {\em et~al.},
\newblock (2007), arxiv:0706.0039.

\bibitem{Akerib:2006ri}
D.~S. Akerib {\em et~al.},
\newblock Nucl. Instrum. Meth. {\bf A559}, 390 (2006).

\bibitem{al.:2007xs}
ZEPLIN-II, G.~J. A.~e. al.,
\newblock (2007), arXiv:0708.1883 [astro-ph].

\bibitem{Bolte:2006pf}
W.~J. Bolte {\em et~al.},
\newblock J. Phys. Conf. Ser. {\bf 39}, 126 (2006).

\bibitem{Baer:2003jb}
H.~Baer, C.~Balazs, A.~Belyaev, and J.~O'Farrill,
\newblock JCAP {\bf 0309}, 007 (2003), hep-ph/0305191.

\bibitem{Dutta:2000hh}
S.~I. Dutta, M.~H. Reno, I.~Sarcevic, and D.~Seckel,
\newblock Phys. Rev. {\bf D63}, 094020 (2001), hep-ph/0012350.

\bibitem{Halzen:2003fi}
F.~Halzen and D.~Hooper,
\newblock JCAP {\bf 0401}, 002 (2004), astro-ph/0310152.

\bibitem{Habig:2001ei}
Super-Kamiokande, A.~Habig,
\newblock (2001), hep-ex/0106024.

\bibitem{Desai:2004pq}
S.~Desai {\em et~al.},
\newblock Phys. Rev. {\bf D70}, 083523 (2004), hep-ex/0404025.

\bibitem{Fogli:2006jk}
G.~L. Fogli, E.~Lisi, A.~Mirizzi, D.~Montanino, and P.~D. Serpico,
\newblock Phys. Rev. {\bf D74}, 093004 (2006), hep-ph/0608321.

\bibitem{Honda:2006qj}
M.~Honda, T.~Kajita, K.~Kasahara, S.~Midorikawa, and T.~Sanuki,
\newblock Phys. Rev. {\bf D75}, 043006 (2007), astro-ph/0611418.

\bibitem{Halzen:2007xx}
F.~Halzen,
\newblock private communication.

\bibitem{Griest:1986yu}
K.~Griest and D.~Seckel,
\newblock Nucl. Phys. {\bf B283}, 681 (1987).

\bibitem{Gould:1987ir}
A.~Gould,
\newblock Astrophys. J. {\bf 321}, 571 (1987).

\bibitem{Winter:2000xx}
K.~Winter,
\newblock Neutrino Physics,
\newblock Cambridge, UK: Univ. Pr. (2000).

\bibitem{Barger:1999fs}
V.~D. Barger, S.~Geer, and K.~Whisnant,
\newblock Phys. Rev. {\bf D61}, 053004 (2000), hep-ph/9906487.

\bibitem{Pasquali:1998xf}
L.~Pasquali and M.~H. Reno,
\newblock Phys. Rev. {\bf D59}, 093003 (1999), hep-ph/9811268.

\bibitem{Bjorken:1979dk}
J.~D. Bjorken and S.~D. Drell,
\newblock RELATIVISTIC QUANTUM FIELD THEORY. (GERMAN TRANSLATION),
\newblock Bibliograph.Inst./mannheim 1967, 409 P.(B.i.-
  hochschultaschenbuecher, Band 101).

\bibitem{Lipari:1993hd}
P.~Lipari,
\newblock Astropart. Phys. {\bf 1}, 195 (1993).

\bibitem{Dutta:2000jv}
S.~I. Dutta, M.~H. Reno, and I.~Sarcevic,
\newblock Phys. Rev. {\bf D62}, 123001 (2000), hep-ph/0005310.

\end{thebibliography}


\end{document}